\documentclass[twocolumn]{aastex631}

\usepackage{amsmath}	
\usepackage{amssymb}	
\usepackage{mathtools}
\usepackage{amsfonts}
\usepackage{graphicx}
\usepackage{appendix}
\usepackage{float}
\usepackage[normalem]{ulem}

\begin{document}

\title{Super-Eddington accretion onto black holes and its application to fallback accretion}
\author{Tamar Faran}\email{tamar.faran@princeton.edu}
\author{Eliot Quataert}
\affiliation{Department of Astrophysical Sciences, Princeton University, Princeton, NJ 08544, USA}
\begin{abstract}
We study the problem of steady-state spherical accretion onto a black hole, in which the internal energy of the flow is governed by radiation and photon diffusion dominates the energy flux at large radii. In the free-fall limit, the fluid equations can admit two types of solutions for a given accretion rate: (1) accretion flows that become isothermal at large radii and (2) solutions in which the temperature at infinity vanishes as a power law of the radius. Using boundary layer theory, we obtain analytic solutions for the two cases and apply our results to fallback accretion onto a black hole following a failed supernova explosion. We give predictions for the observational signature of fallback accretion using realistic progenitor properties from MESA, both for a fully ionized inflow and for the more realistic case in which recombination/ionization take place due to low photospheric temperatures. The observed fading sources coincident with the failed-supernova candidates in NGC 6946 and M31 are too luminous to be powered by spherical accretion onto newly formed black holes; the observed sources are instead likely due to accretion of the turbulent, convective envelope of the supergiant progenitor.
\end{abstract}

\section{Introduction}
Accretion of matter onto a black hole can in principle occur at accretion rates well in excess of the Eddington limit, $\dot{M}>L_E/c^2$, where $L_E$ is the Eddington luminosity
\begin{equation}
    L_E = \frac{4\pi c\, G M}{\kappa}\,,
\end{equation}
$c$ is the speed of light, $G$ is the gravitational constant, $M$ is the mass of the black hole, and $\kappa$ is Thomson's opacity. Such high accretion rates are expected to generate diffusive luminosities that can significantly affect the dynamics of the inflowing material, contrary to adiabatic accretion \citep[see e.g.,][]{Begelman1978,Flammang1982,Flammang1984,Vitello1984}.   

One scenario in which super-Eddington accretion rates should generically occur is fallback accretion onto newly formed black holes in core-collapse supernovae.   In the standard picture of core-collapse supernova explosions, a strong shock wave is launched by neutrino energy deposition in the seconds following the bounce of the iron core, with enough energy to unbind most of the stellar envelope \citep{Burrows2021}; this leaves behind a neutron star remnant or, in some cases, a black hole. Both neutron star and black hole formation in successful supernovae can lead to super-Eddington fallback accretion because not all of the stellar envelope is unbound.   
However, it is also plausible that some fraction, perhaps $\sim 10\%$, of massive star core-collapse  do not result in a successful explosion \citep[e.g.,][]{Ertl2016,Sukhbold2016}; instead, a significant fraction of the envelope remains bound and collapses to form a black hole. This scenario is sometimes referred to as a `failed supernova', and is supported by the observational deficiency of supernova progenitors with masses greater than $\sim 20 \text{M}_\odot$, the `red supergiant problem' (e.g., \citealt{Smartt2009,Smartt2015,Davies2018}; see, however, \citealt{Beasor2025}).   There are also two candidate failed SNe.   In these events, a massive red/yellow supergiant progenitor star `disappeared' without any accompanying energetic explosion \citep[][]{Kochanek2024,De2024}, although with signatures of the weak transient expected from collapse to a black hole \citep[][]{Lovegrove2013,Lovegrove2017,Antoni2023}.


Motivated by the application to fallback accretion in supernovae, this paper revisits the problem of super-Eddington accretion using a fully analytic approach. We apply our results to fallback accretion from failed supernovae to constrain properties of these events.   In our analysis we focus on spherical accretion, thereby neglecting energy dissipation via rotation, turbulence, etc; we return to the limits of this approximation in \S \ref{sec:Summary}.


The structure of this paper is as follows.
In \S \ref{sec:FluidEquations} we introduce the fluid equations and derive their dimensionless form. In \S \ref{sec:BifurcatingSolutions} we show that the equations admit two separate branches of solutions and obtain analytic solutions for the two cases in \S \ref{sec:BoundaryLayerAnalysis}. We apply our solutions to fallback accretion in \S \ref{sec:Applications} and summarize in \S \ref{sec:Summary}.   Readers interested primarily in the application to  newly formed black holes  in massive star collapse can focus on \S \ref{sec:Applications}.

\section{Governing equations}\label{sec:FluidEquations}
We solve for the dynamics of a steady accretion flow onto a black hole under the following assumptions and simplifications. We begin by taking the limit in which the energy density in photons dominates over gas thermal energy; the adiabatic index of the flow is therefore constant and is equal to $\gamma = 4/3$. We initially assume that the gas is fully ionized, and that the opacity $\kappa$ is given by Thomson scattering; we relax this assumption in \S \ref{sec:recombination} to account for the change in opacity due to recombination.  We neglect rotation and energy dissipation due to turbulence (e.g., convection in a RSG envelope). Nevertheless, the flow is not isentropic, as the outer boundary of our domain is at an optical depth of unity, and photons from inner regions can diffuse to large radii over times shorter than the dynamical time of the system.
Finally, we work under the assumption that the self-gravity of the accreting material is negligible compared to that of the black hole, so that the gravitational potential is that of a point source. Using the Paczyński–Wiita potential, the stationary Newtonian fluid equations in spherical symmetry, neglecting the contribution of gas pressure, take the following form:
\begin{subequations}\label{eq:FluidEqns}
    \begin{equation}\label{eq:Mdot}
        A(r) \rho v = \dot{M} = const
    \end{equation}
    \begin{equation}\label{eq:MomentumEqn}
        p'/\rho = -(v^2)'/2-G M/(r-r_S)^2
    \end{equation}
    \begin{equation}\label{eq:EnergyEqn}
       - \frac{1}{\gamma-1}\left(p/\rho\right)'-p\left(1/\rho\right)' = -\nabla \cdot \mathcal{F}/(\rho v)
    \end{equation}
\end{subequations}
where $\rho$ is the material density, $A(r) = 4 \pi r^2$ is the cross section, $r_S$ is Schwarzschild's radius
\begin{equation}
    r_S\equiv\frac{2 G M}{c^2}\,,
\end{equation}
$\mathcal{F}$ is the diffusive flux under the diffusion approximation:
\begin{equation}
   \mathcal{F} = -\frac{c}{\kappa}\frac{p'}{\rho} \,,
\end{equation}
and is related to the diffusive luminosity by $L_d = A(r) \mathcal{F}$. The velocity $v$ and the accretion rate $\dot{M}$ are both defined positive.
Equations \eqref{eq:EnergyEqn} and \eqref{eq:MomentumEqn} can be written as $-\dot{M} Be' = -L_d'$, and upon integration one obtains the energy conservation relation
\begin{equation}\label{eq:Bernoulli}
    \dot{E} = L_d-\dot{M}\left(\frac{v^2}{2}-\frac{G M}{r-r_S}+\frac{\gamma}{\gamma-1}\frac{p}{\rho}\right)\,,
\end{equation}
where $\dot{E}$ is a constant of integration, determined by the boundary conditions of the system.
\subsection{Dimensionless variables}
For the clarity of the analysis, we will hereafter work with dimensionless variables. Let us first introduce the variable $\epsilon$ that measures the deviation of the flow from the free-fall limit, and is related to the velocity through
\begin{equation}\label{eq:epsilon_def}
    v^2 = \frac{2 G M}{r-r_S}(1-\epsilon)\,.
\end{equation}
$\epsilon=0,1$ in the limits where the flow is exactly in free fall or at rest, respectively. Second, we define the dimensionless accretion rate as
\begin{equation}
    \dot{m} \equiv \frac{\dot{M} c^2}{L_E}\,,
    \label{eq:mdotEdd}
\end{equation}
where $\dot{m} \gg 1$ for the case of super-Eddington accretion. We will also make use of the following definitions:
\begin{equation}\label{eq:DimensionlessTransform}
\begin{split}
    &\xi \equiv \frac{r}{r_S}\,;~~ \Lambda_d \equiv \frac{L_d}{\dot{M}c^2}\,; ~~ \Pi \equiv \frac{p}{\rho(r_S) c^2} \,;~~\\& \Psi \equiv \frac{\epsilon}{\xi -1}\,;~~ \dot{e}\equiv\frac{\dot{E}}{\dot{M}c^2};~~ \Lambda_a \equiv \frac{\dot{M}(4 p/\rho)}{\dot{M} c^2}\,.
\end{split}
\end{equation}
The dimensionless pressure can be written explicitly in terms of fundamental constants:
\begin{equation}
    \Pi = \frac{16 \pi^3 \alpha^3}{135}\frac{m_e}{m_p}\frac{r_S}{r_e}\left(\frac{k_B T}{m_e c^2}\right)^4 \dot{m}^{-1} \simeq \frac{7\times 10^{-14} M_{10} \,T_4 ^4}{\dot{m}}\,,
\end{equation}
where $\alpha$ is the fine structure constant, $r_e$ is the classical electron radius,  $m_e$ and $m_p$ are the electron and proton mass, respectively, $k_B$ is Boltzmann's constant and $T$ is the temperature. We denote $T_4 \equiv T/10^4\text{K}$ and $M_{10} = M/10 M_\odot$. Eq \eqref{eq:Bernoulli} can now be rewritten in terms of the dimensionless quantities:
\begin{equation}\label{eq:BernoulliDimensionless}
    \dot{e} =\frac{\Psi}{2}+ \Lambda_d - \Lambda_a\,.
\end{equation}
$\Psi$ can be interpreted as the local rate at which internal energy is added to the flow due to compression.
Since we are interested in systems that become diffusive to photons at large radii, it is also useful to define the photospheric radius, where $\tau \equiv \int \kappa \rho dr = 1$:
\begin{equation}
\xi_{ph} = \frac{r_{ph}}{r_S} = \frac{f_\tau \dot{m}^2}{1-\epsilon_{ph}} \,.
\end{equation}
The constant $f_\tau$ depends on how the density varies with the radius. In the free-fall limit $f_\tau = 1$.

\subsection{Dimensionless equations}
The spatial variable $\xi$ ranges between $1$ and $\infty$. However, in the following analysis it will be more convenient to work in a finite domain. For that purpose, we make a transformation to a new independent variable
\begin{equation}\label{eq:x_def}
    x \equiv1-\xi^{-1}~;~x \in \{0, 1\}\,.
\end{equation}
In terms of $x$, there exists a simple relation between $\Psi$ and $\Lambda_d$:
\begin{equation}\label{eq:Lambdad_PsiPrime}
   \Lambda_d   = -\delta\,\Psi'(x)\,,
\end{equation}
where we defined the small parameter
\begin{equation}
    \delta \equiv \dot{m}^{-1}\,.
\end{equation}
Since $\dot{m}\gg1$ in super-Eddington flows, $\delta \ll 1$, a fact that will become useful later in the context of boundary layer theory.
Using equations \eqref{eq:FluidEqns},  \eqref{eq:DimensionlessTransform} and \eqref{eq:x_def}, we obtain a single second order ordinary differential equation (ODE) for $\Psi(x)$:
\begin{multline}\label{eq:PsiODE}
       \delta\Psi''F_1[x,\Psi]+\Psi' F_2[x,\Psi] -2 \delta (\Psi')^2 x^2(1-x)+ \\ \Psi\left[1-5x+4x^2(1-2\dot{e})\right] + 4 x^2 \Psi^2 \\ -2\dot{e}(1-x)(1-4x) = 0
\end{multline}
where
\begin{multline*}
    F_2[x,\Psi] = -2(1-x)(3x-3x^2+\dot{e} x -4\delta x +\delta)\\+ x^2(7-7x-8\delta)\Psi
\end{multline*}
and
\begin{equation*}
    F_1[x,\Psi] = 4x(1-x)\left[x\Psi-(1-x)\right]\,.
\end{equation*}
Eq \eqref{eq:PsiODE} is non-linear in $\Psi$, and as a result, difficult to solve analytically. Fortunately, Eq \eqref{eq:PsiODE} can be simplified when the analysis is restricted to the limit in which the flow is very close to free-fall. At the photosphere, $\Psi \sim \epsilon/\xi \sim \epsilon \delta^2 \ll 1$ and near the black hole $\Psi \sim \epsilon \sim 0$. Since there are no energy sources in the system, $\dot{e}$ is maximal when the energy flux is governed by local compressional work, in which case $\mathcal{O}(\dot{e}) =\Psi$ (see Appendix \ref{app:edoEvaluation}). This motivates the linearization of \eqref{eq:PsiODE} by eliminating the terms $\Psi^2$, $\Psi'^2$, $\Psi\, \dot{e}$ and $\Psi'\, \dot{e}$:
\begin{multline}\label{eq:PsiODELinearized}
    4x(1-x) \delta \Psi''+\left[6x(1-x)+2(1-4x)\delta\right] \Psi' \\ -(1-4x) (\Psi-2\dot{e}) = 0\,.
\end{multline}
Now, letting
\begin{equation}\label{eq:ChiDefinition}
    \chi \equiv \Psi-2\dot{e} = 2(\Lambda_a-\Lambda_d)\,,
\end{equation}
we obtain
\begin{multline}\label{eq:ChiODELinearized}
    4x(1-x) \delta \chi''+\left[6x(1-x)+2(1-4x)\delta\right] \chi' \\ -(1-4x)\chi
    = 0\,.
\end{multline}
The parameter $\dot{e}$ does not appear explicitly in Eq \eqref{eq:ChiODELinearized}, while Eq \eqref{eq:PsiODE} contains $\dot{e}$ even when it is written in terms of $\chi$. In the full, nonlinear problem, that extends to where $\mathcal{O}(\epsilon) \sim 1$, $\dot{e}$ is an eigenvalue and is determined by smoothly integrating the equation across a movable singularity that appears in the vicinity of the adiabatic sonic point. The singularity is not encountered when the photosphere is in the free-falling region, as the former lies outside of the solution's domain. The value of $\dot{e}$ in this case is determined by the outer boundary conditions. The linearized equation \eqref{eq:ChiODELinearized} retains all of the relevant dynamics in the free-fall limit ($\epsilon \ll 1$) and its solution is indistinguishable from that of the nonlinear equation \eqref{eq:PsiODE} in that regime.

The new variable $\chi$ is essentially twice the difference between the advective and diffusive luminosities. Since the flow is assumed to be diffusive to photons at the outer boundary, $\chi$ has to vanish somewhere in the domain: it acquires positive values at small radii and negative values at large radii.

\subsection{Numerical integration}\label{sec:NumericalIntegration}
Integration of Eq \eqref{eq:ChiODELinearized} requires two boundary conditions. Near the horizon of the black hole, $\mathcal{F} = 0$, and we must require $\chi' = 0$ there. The exact value of $x$ where this condition is enforced is not important for the properties of the flow at large radii, and it is sufficient to require that the solution is consistent with $\chi'(x\rightarrow0) = 0$. If the exact details of the flow at small radii were important, one would have to solve the general relativistic equations. The second boundary condition is imposed on the value of $\chi$ at the outer or inner boundary, and can be associated with a boundary condition on the temperature.

\begin{figure}
    \includegraphics[width=0.85\linewidth]{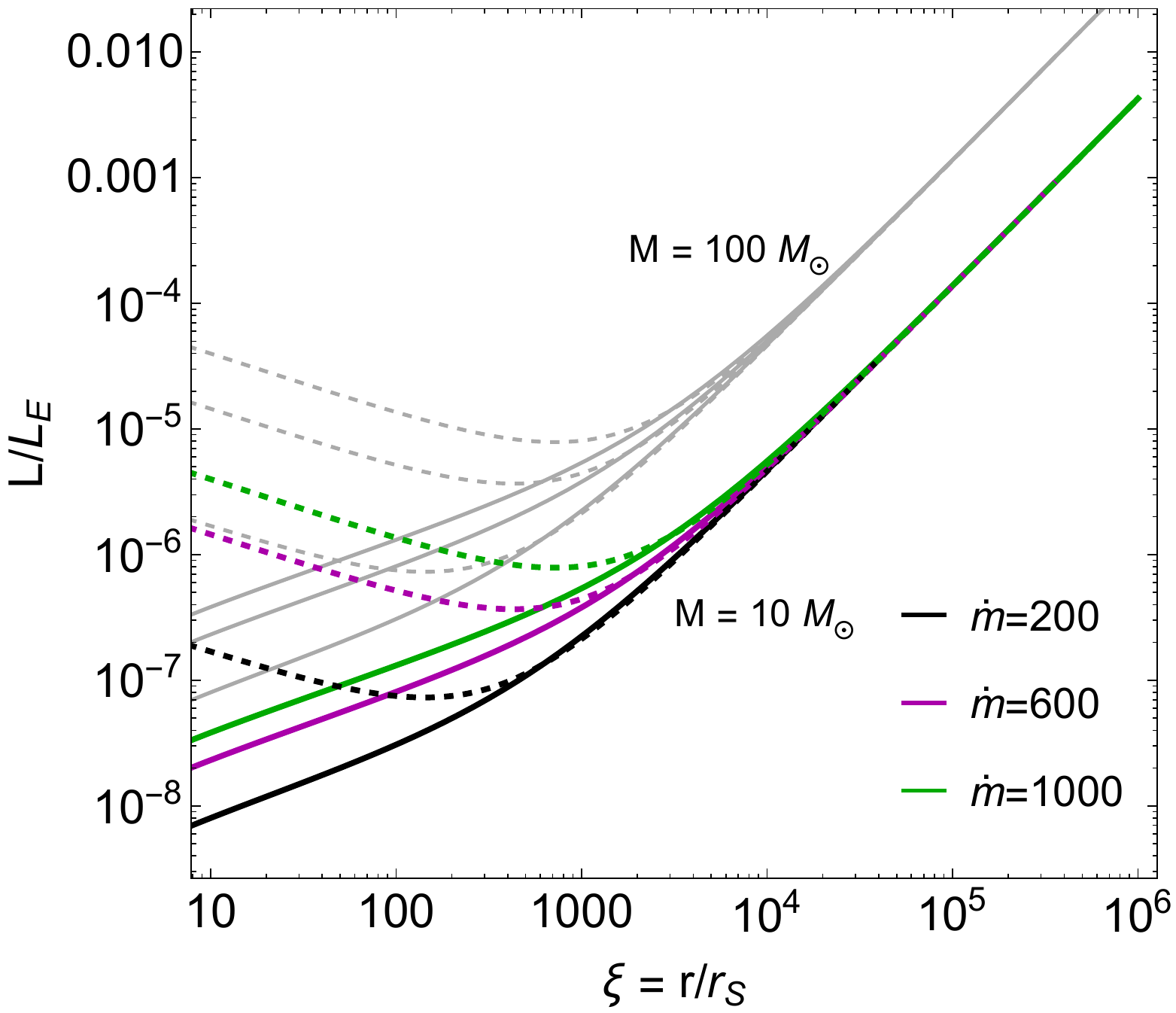}
    \includegraphics[width=0.85\linewidth]{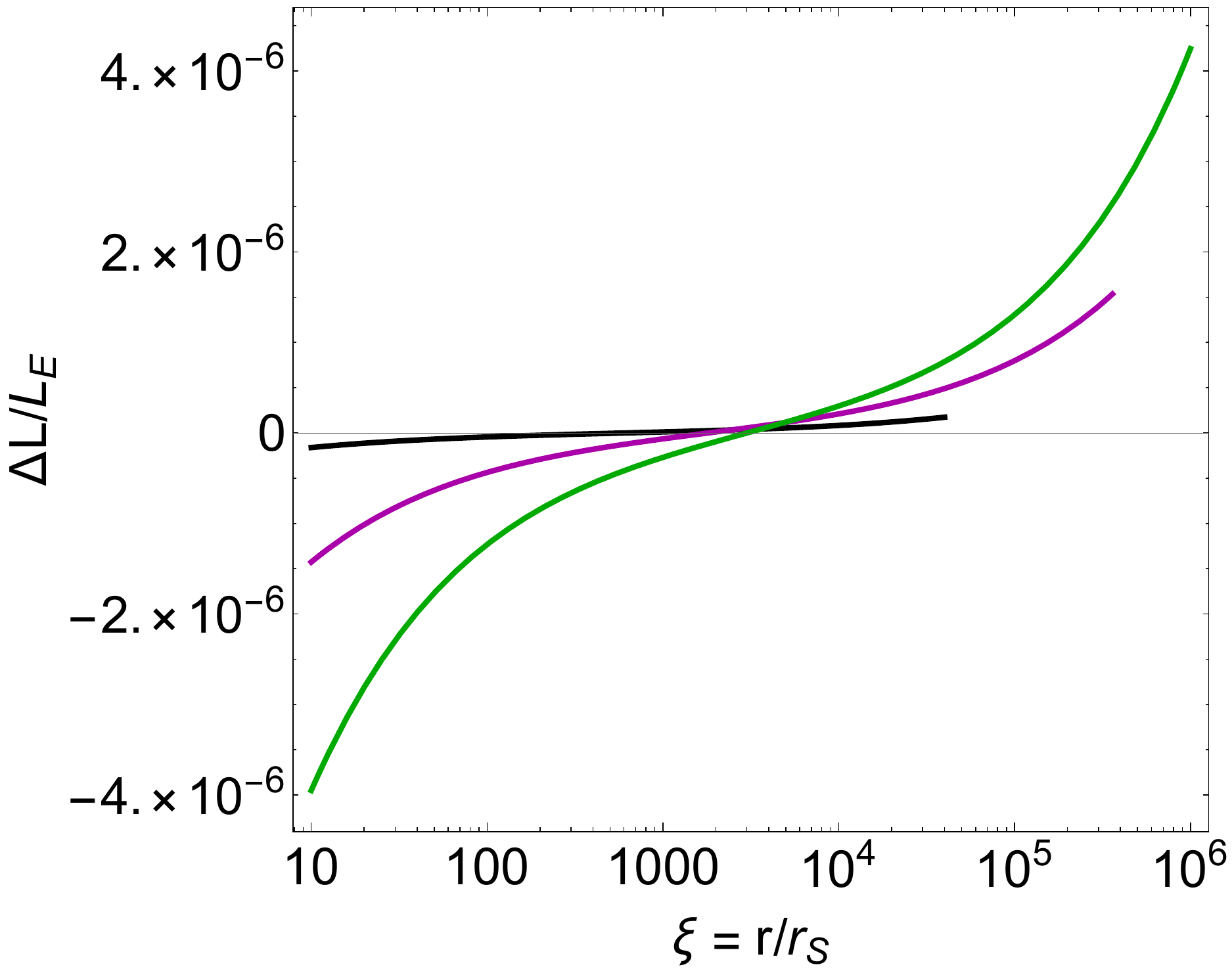}
    \includegraphics[width=0.85\linewidth]{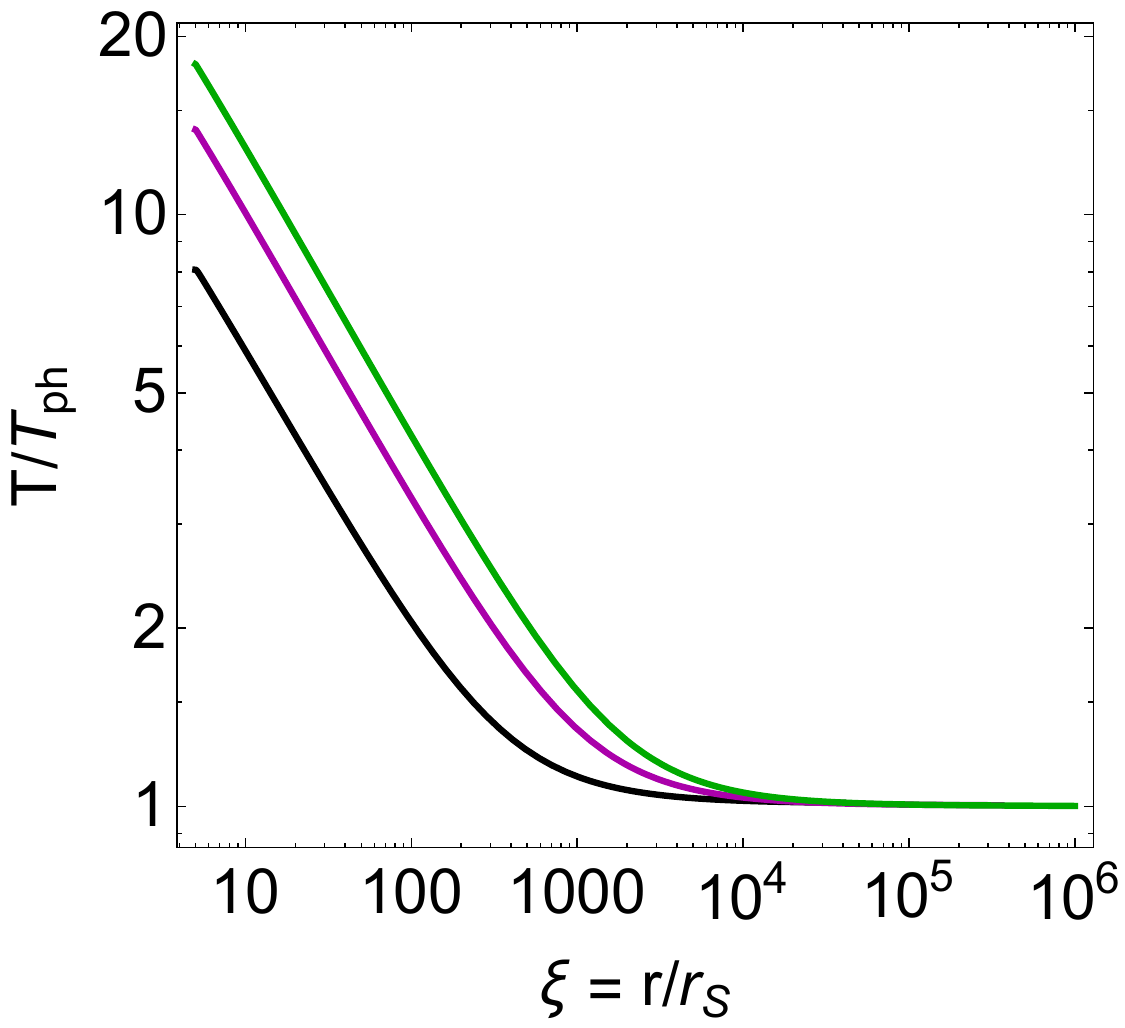}
    \caption{Diffusive luminosity (solid lines) and advective luminosity (dashed lines) for a range of $\dot{m}$ values and black hole masses of $10 M_\odot$ and $100 M_\odot$ (top panel), the rest frame luminosity (middle panel) and the corresponding temperature, independent of the mass of the black hole (bottom panel). The boundary conditions are $L_d(r\rightarrow0) = 0$ and a photospheric temperature $T_{ph} = 10^4$ K.}
    \label{fig:SolVarmdot}
\end{figure}

\begin{figure}
    \includegraphics[width=0.9\linewidth]{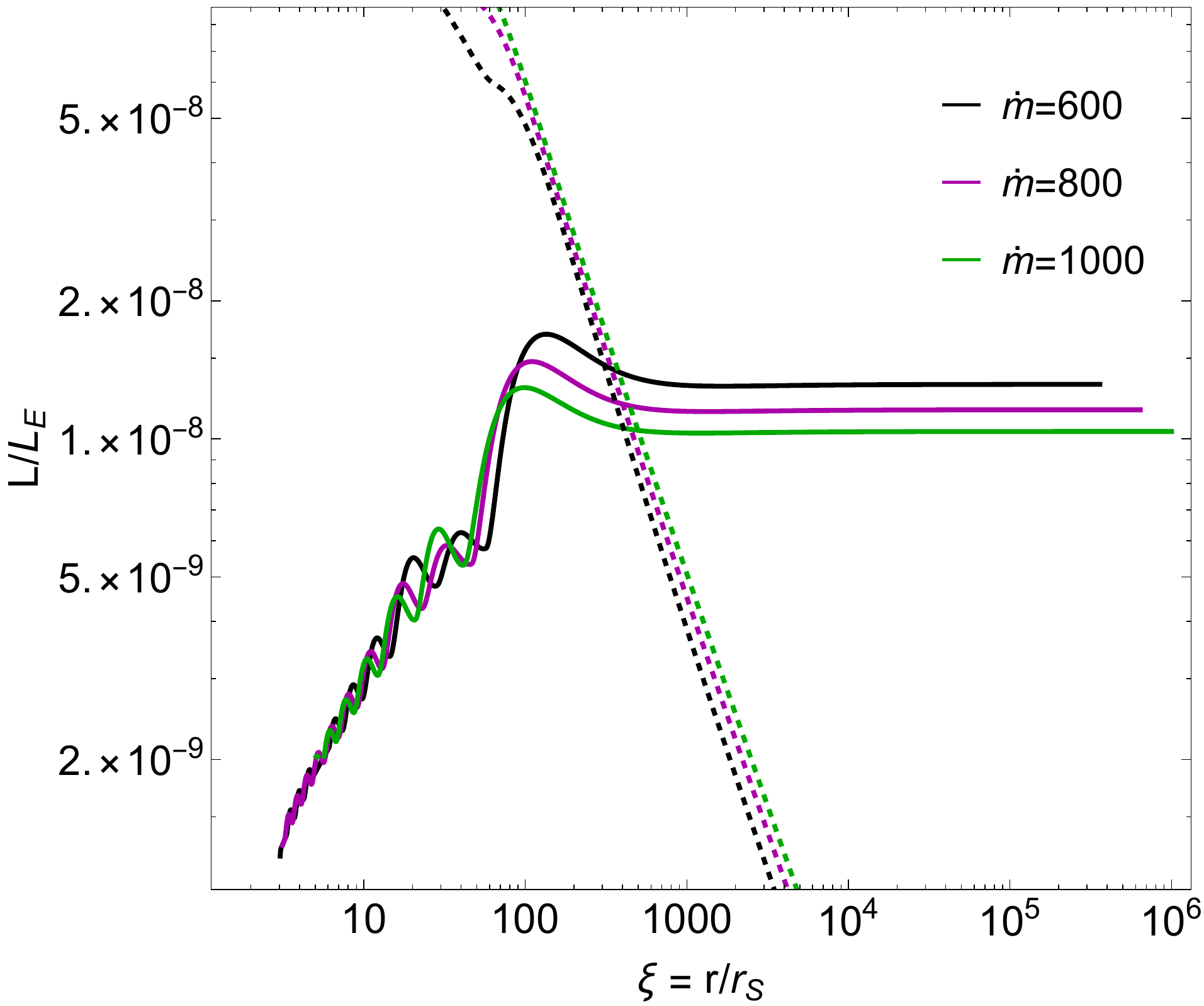}
    \includegraphics[width=0.9\linewidth]{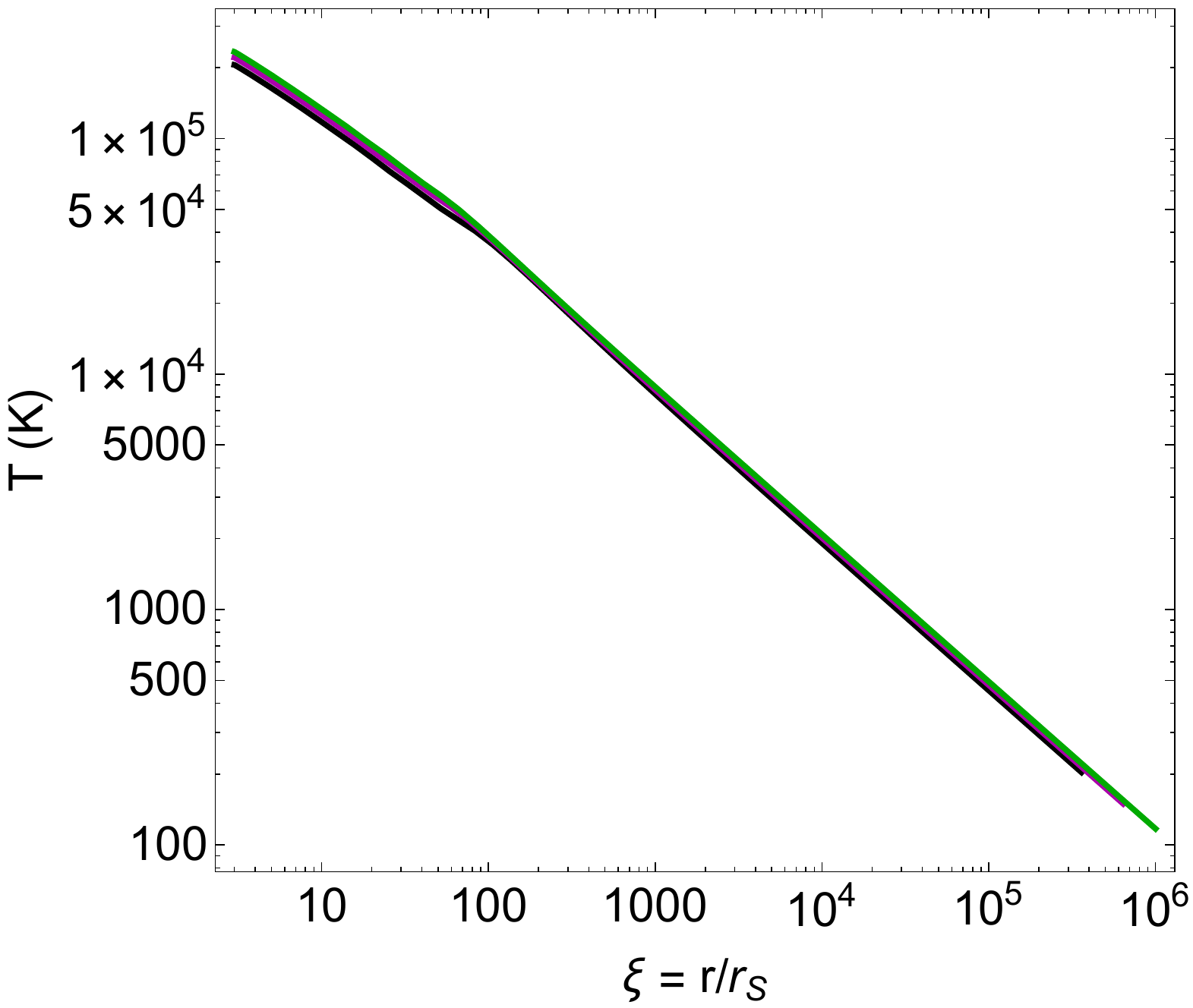}
    \caption{Same as Fig \ref{fig:SolVarmdot} for the powerlaw solution branch. Here the rest-frame luminosity is to high accuracy equal to the diffusive luminosity and is independent of the black hole mass. The boundary condition was imposed on the value of $\chi$ at the inner boundary in addition to $\chi' \rightarrow0$ there.}
    \label{fig:SolVarmdotPL}
\end{figure}

Numerical integration of Eq \eqref{eq:ChiODELinearized}, or equivalently, Eq \eqref{eq:PsiODE}, reveals that they bifurcate to two different branches of solutions for a given value of $\dot{m}$: one, in which the pressure at large radii asymptotes to a constant (Fig \ref{fig:SolVarmdot}), and another, in which the pressure vanishes at infinity as a power-law of the radius (Fig \ref{fig:SolVarmdotPL}). The properties of the two branches are analyzed analytically in the following sections.

\section{Bifurcation analysis}\label{sec:BifurcatingSolutions}
Below, we present a qualitative analysis of the bifurcating nature of Eq \eqref{eq:ChiODELinearized} that is the origin of the two types of solutions shown in Fig \ref{fig:SolVarmdot}-\ref{fig:SolVarmdotPL}.

\subsection{Bifurcation analysis}
\begin{figure}
    \centering
    \includegraphics[width=0.95\columnwidth]{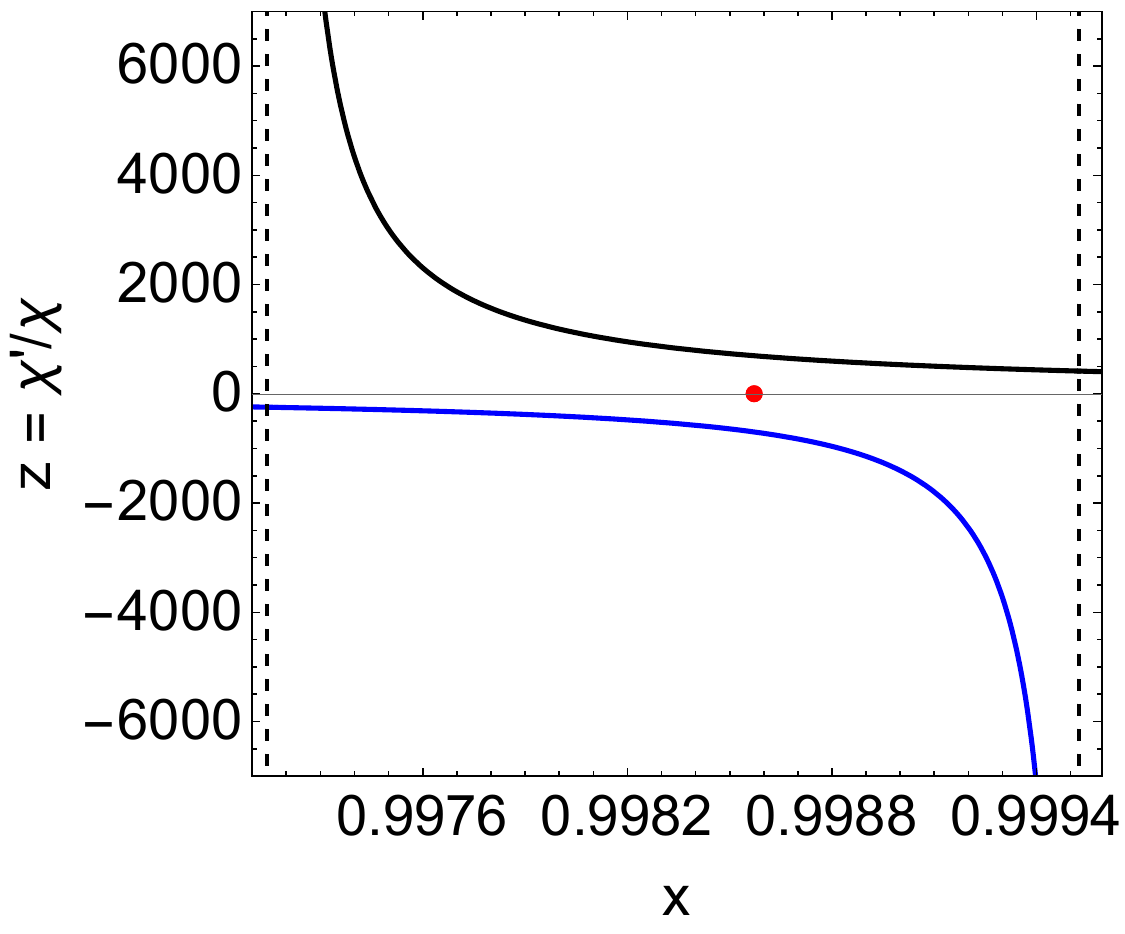}
    \caption{Bifurcating solutions around $x_{bf} \simeq 1-\delta$, designated by the red point. The curves can cross the $x$-axis only through infinity at the positions marked by the dashed lines.}
    \label{fig:z_bifurcation}
\end{figure}

Eq \eqref{eq:ChiODELinearized} is of the form
\begin{equation}\label{eq:eq_chi}
    \chi''+a(x) \, \chi' + b(x) \chi = 0\,.
\end{equation}
The equation is equidimensional in $\chi$, and its order can be reduced via the transformation $\chi = e^u$ and $u' = z$, resulting in a Riccati equation:
\begin{equation}\label{eq:BifurcationEq1}
    z' = -z^2 - a(x) z -b(x)\,,
\end{equation}
where
\begin{equation}\label{eq:z_def}
    z = \frac{\chi'}{\chi}\,.
\end{equation}
We now rewrite Eq \eqref{eq:BifurcationEq1} as
\begin{equation}\label{eq:BifurcationEq2}
    z' = -\left[z+a(x)/2\right]^2 + a(x)^2/4 -b(x)\,.
\end{equation}
The first and third terms on the right-hand side are always negative, and $b(x)>a(x)^2/4$ in the region $1-3 \delta\lesssim x\lesssim 1-\delta/3$. Since $a(x)=0$ at 
\begin{equation}\label{eq:x_bf}
    x_{bf} = \frac{1}{6}\left(3-4\delta + \sqrt{9+4\delta(4\delta-3)}\right) = 1-\delta + \mathcal{O}(\delta^2)\,,
\end{equation}
the right-hand side of Eq \eqref{eq:BifurcationEq2} is negative for several (radial) scale heights around the bifurcation point, $x_{bf}$. According to Eq \eqref{eq:Lambdad_PsiPrime}, $\chi' = \Psi' <0$ since the diffusive luminosity ($-\Psi'$) cannot be negative, and the equation admits different dynamics depending on the sign of $\chi$, and therefore $z$, around $x_{bf}$, as shown in Fig \ref{fig:z_bifurcation}. In other words, the behaviour of the flow is set by whether the diffusive luminosity is greater or smaller than the advective luminosity at $x_{bf}$. The main characteristics of the two solutions are investigated below. \newline

\noindent{\bf Case $z_{bf}>0$}:\newline
In this branch, $\chi = 2(\Lambda_a-\Lambda_d)<0$ around $x_{bf}$, which implies that the diffusive luminosity is greater than the advective luminosity. $z$ remains positive as $x$ increases, although $z'<0$ throughout the region; $z$ therefore must asymptote to a constant. Very close to $x=1$, $z' \simeq 0$ and Eq \eqref{eq:BifurcationEq1} becomes an algebraic equation for $z$:
\begin{equation}\label{eq:z_near_xs}
    z \simeq \frac{-a(x) \pm \sqrt{a(x)^2 -4 b(x)}}{2}\,.
\end{equation}
The negative root diverges at $x\rightarrow 1$ and the solution is given by the positive root. In the limit $x\rightarrow 1$:
\begin{equation}
    z(x\rightarrow1) =\frac{\chi'}{\chi}\bigg|_{x\rightarrow1} = \frac{1}{2\delta}\,.
\end{equation}
This solution describes systems with vanishing pressure at infinity, which is inferred from the fact that the diffusive luminosity becomes constant. $z$ diverges where $\chi=0$, at a singular point denoted by $x_s$. Clearly, $z$ is very large near $x_s$, so that the last two terms in Eq \eqref{eq:BifurcationEq1} can be neglected to give $z' \simeq -z^2$, which is solved by
\begin{equation}\label{eq:z_pole}
    z = \frac{1}{x-x_s}\,.
\end{equation}
Using Eq \eqref{eq:z_def} to solve for $\chi$, we obtain
\begin{equation}\label{eq:ChiConstLd}
    \chi = C(x-x_s)\,,
\end{equation}
where $C$ is a constant, equal to $-L_\infty/L_E$, where $L_\infty$ is the asymptotic diffusive luminosity at infinity. Since $\Lambda_d = -\delta\chi'$, this solution means nothing but constant diffusive luminosity at large radii.
The location of $x_s$ can be estimated to leading order by matching Eq \eqref{eq:z_pole} with the asymptotic solution at $x=1$, which gives
\begin{equation}
    x_s = 1-2\delta\,.
\end{equation}
The point at which $\chi=0$, or $L_d=L_a$, is usually referred to as the `trapping radius', beyond which photons can escape the system at times shorter than the dynamical time. In terms of the radial coordinate, the trapping radius is located at 
\begin{equation}
    r_{tr} = (\dot{m}/2) \times r_S\,.
\end{equation}
It should be noted that in this branch, the flow never becomes sub-sonic below the photosphere in the free-fall limit: beyond the trapping radius, which lies in the supersonic region, the velocity, the escape velocity and the adiabatic sound speed all scale as $r^{-1}$, and can never intersect with each other.
\newline

\noindent{\bf Case $z_{bf}<0$}:\newline
In this branch, $\chi= 2(\Lambda_a-\Lambda_d)>0$ around $x_{bf}$ and the net outgoing luminosity is still negative there. $z$ becomes more and more negative as $x$ increases. Once $\chi=0$, $z$ becomes infinite and crosses 0. In the limit $x\rightarrow1$, $a(x)$ is negative and exceeds $b(x)$. The asymptotic solution in this limit is more easily derived in terms of $\chi$. Approximating Eq \eqref{eq:eq_chi} as $\chi''/\chi' = -a(x)$, we can immediately integrate and obtain
\begin{equation}\label{eq:ChiPrimeNegzEstimate}
    \chi' \sim C \, \frac{e^{-\frac{3 x}{2\delta}}}{[x(1-x)]^{3/2}}\,,
\end{equation}
so that $z$ diverges at $x\rightarrow1$ ($C$ is some constant). Since $\chi$ is finite, the diffusive and net luminosities diverge like $(1-x)^{3/2} \propto r^{3/2}$. As we shall see in \S \ref{sec:BLAnalysisIsothermal}, the luminosity is a result of isothermal compression. Outside the trapping radius, $c_s^2 = \gamma \,p/\rho \propto r^{3/2}$ and $v^2 \propto r^{-1}$. Since there are no additional velocity scales in the system, the Mach number decreases with radius and the flow may become subsonic either slightly below or above the photosphere.\newline

Both types of solutions are stable against convection, as the flow is in general super-sonic in all or most of the domain. Even if the solutions admit regions of negative entropy gradients, subsonic buoyant motions will be advected inwards with the flow.\newline

A key property of Eq \eqref{eq:ChiODELinearized} is that it can admit boundary layers. Depending on the desired boundary conditions, $\chi$ may be required to change very rapidly in a narrow region in order to satisfy the boundary conditions. The plausibility of forming a boundary layer relies upon the fact that the highest derivative of $\chi$ is multiplied by the small parameter, $\delta$; while $\chi$ itself is not necessarily large, its derivatives can be large since $\chi''$ is multiplied by $\delta\ll1$. According to boundary layer theory \citep[e.g.,][]{BenderOrszag}, boundary layers can form either at the boundaries of the domain (i.e., at $x=1$ or $x=0$), or at the point where $a(x)=0$, i.e., at $x_{bf}\simeq 1-\delta$. Since $a(x)<0$ at $x>x_{bf}$, a boundary layer can exist at $x=1$, but not at $x=0$. In \S \ref{sec:BoundaryLayerAnalysis} we employ boundary layer theory to obtain solutions for $\chi$ in the two branches. We will show that the negative branch forms a boundary layer at $x=1$ and the positive branch forms one at $x = x_{bf}$.

\section{Boundary Layer Solutions}\label{sec:BoundaryLayerAnalysis}
In the following, we use boundary layer analysis to obtain approximate analytic solutions to Eq \eqref{eq:ChiODELinearized}.
\subsection{Isothermal solutions, $z_{bf}<0$}\label{sec:BLAnalysisIsothermal}
Let us consider the possibility that a boundary layer forms near $x=1$. This is a natural guess for the negative branch due to the fact that $\chi'$ starts increasing more rapidly outside the trapping radius. A dominant balance analysis suggests that the thickness of the region where $\chi$ changes rapidly is of order $\delta$. We introduce the auxiliary variable within the boundary layer:
\begin{equation}\label{eq:defXBLx1}
    X \equiv \frac{1-x}{\delta}\,~; ~ \chi(x) = Y(X)\,.
\end{equation}
$X$ changes from $0$ at $x=1$ to $\infty$ at $x=0$.
We will look for a solution in the form of a perturbation series
\begin{equation}\label{eq:YSeriesBLx1}
    Y = \sum _{n=0}^{\infty}Y_n \cdot \delta^n\,.
\end{equation}
Substituting Eq \eqref{eq:YSeriesBLx1} into Eq \eqref{eq:ChiODELinearized} and collecting orders of $\delta$, we obtain the following ODEs:
\begin{multline}\label{eq:ODESystemBLx1}
    4X\,Y_n''+6(1-X)Y_n'+3Y_n \\=
    \begin{cases}
        0 &\,, n=0 \\4X^2 Y_{n-1}''+2X(4-3X)Y_{n-1}'+4X Y_{n-1}&\,, n>0
    \end{cases}\,.
\end{multline}
The leading order solution for $n=0$ is
\begin{equation}
    Y_0 =C_1\sqrt{\frac{3 X}{2}}\left(1-\frac{1}{3X}\right) +C_2L_{1/2}^{1/2}(3X/2) \,,
\end{equation}
where $L_{1/2}^{1/2}(3X/2)$ is the generalized Laguerre polynomial\footnote{ $L_{1/2}^{1/2}\left(\frac{3 X}{2}\right)=  \frac{2}{\pi}e^{3X/2}- \sqrt{\frac{2}{3\pi}}\frac{1-3X}{\sqrt{X}}\int_0^{\sqrt{\frac{3X}{2}}} e^{y^2}dy$}.
To avoid divergence at $X\rightarrow\infty$ we must set $C_2 = 0$. Changing back to $x$, we are left with
\begin{equation}\label{eq:chi0_BLx1}
    \chi_0 = C_1 \sqrt{\frac{3 (1-x)}{2\delta}}\left[1-\frac{\delta}{3(1-x)}\right]\,.
\end{equation}
To leading order, we see that $\chi=0$ at $x = 1-\delta/3$, which defines the trapping radius in this branch of solutions:
\begin{equation}
    r_{tr} = 3\dot{m}\times r_S\,.
\end{equation}
From the relation between the diffusive luminosity and the rest-frame luminosity, we find that $C_1 \simeq 1.3 \times 10^{-12}~ \dot{m}^{1/2} M_{10} T_{\infty,4}^4$, where $T_{\infty}$ denotes the asymptotic temperature . 
In physical units, Eq \eqref{eq:chi0_BLx1} reads
\begin{equation}\label{eq:DelL_r_isothermal}
    \Delta L(r) = \frac{\frac{r/r_S}{3\dot{m}}-1}{\sqrt{r/r_S}} ~(1.6\times 10^{-12}~ \dot{m}^2M_{10} T_{\infty,4}^4 )\times L_E\,,
\end{equation}
where $\Delta L = L_d-L_a$ is the rest frame luminosity. The diffusive and advective luminosities are effectively equal beyond the trapping radius, as demonstrated in Fig \ref{fig:IsothermalLum}. This property of radiation dominated accretion flows was previously identified in numerical simulations by \cite{Vitello1984}.

\begin{figure}
    \centering
    \includegraphics[width=0.98\linewidth]{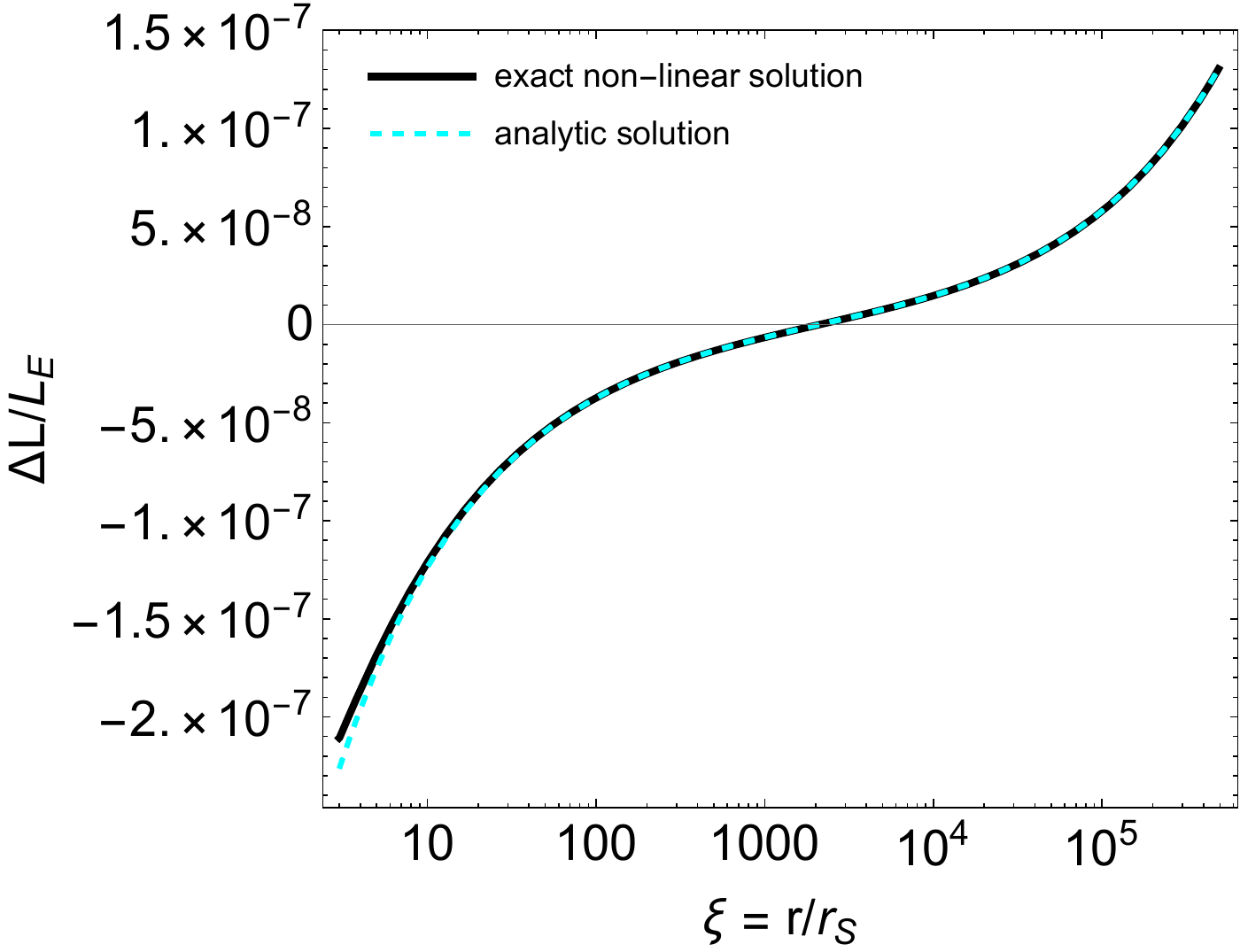}
    \includegraphics[width=0.98\linewidth]{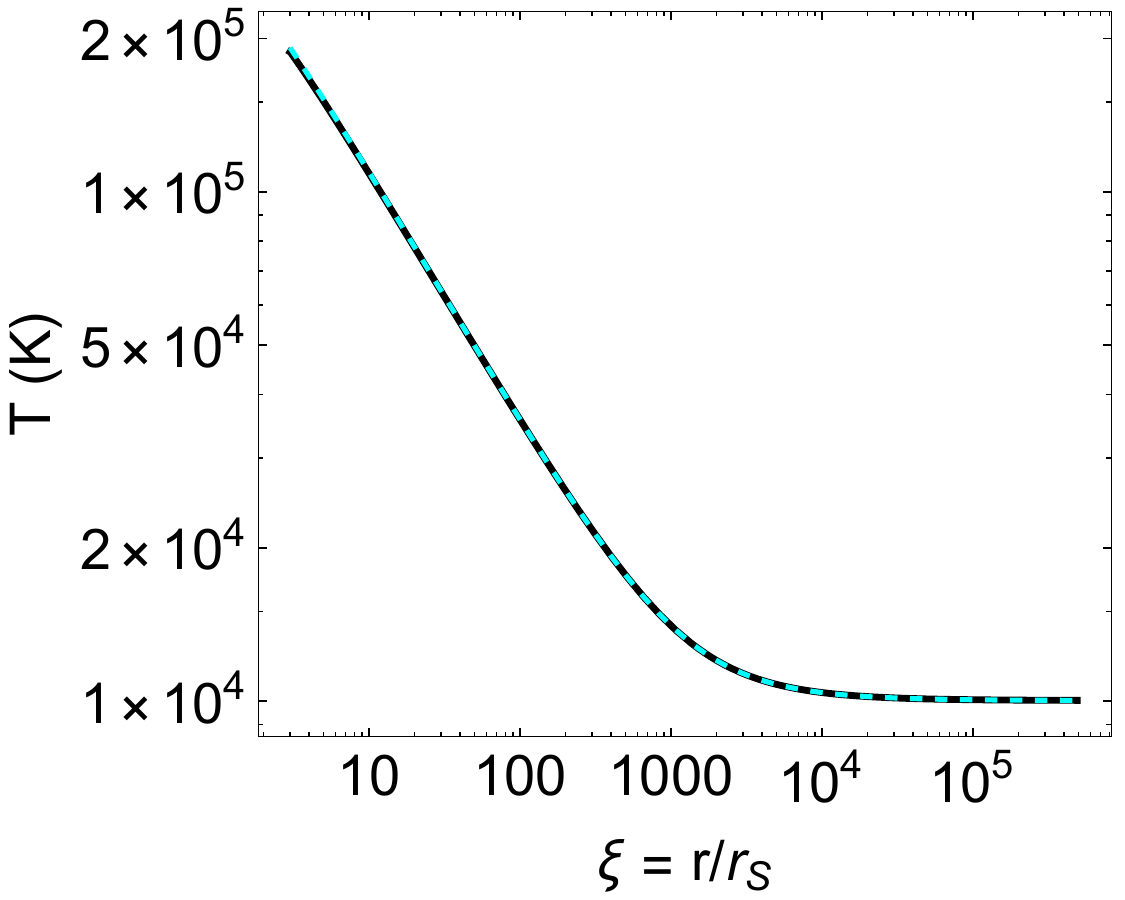}
    \caption{Analytic and non-linear numerical solution for the isothermal branch, with parameters $\dot{m} = 700$, $M = 10 M_\odot$ and a photospheric temperature of $T = 10^4$K.}
     \label{fig:IsothermalSolutions}
\end{figure}

The pressure can be derived from equations \eqref{eq:BernoulliDimensionless} and \eqref{eq:chi0_BLx1}:
\begin{multline}
    \Pi \simeq \frac{(1-x)^{3/2}}{8}\left(\chi-\delta \chi'\right) \\= \frac{3}{16}\sqrt{\frac{\delta}{6}}\left[(1-x+\delta/3)+\frac{2(1-x)}{\delta}(1-x-\delta/3)\right]C_1\,.
\end{multline}
In physical units,
\begin{equation}
    p(r) = \frac{\frac{r/r_S}{\dot{m}}\left(1+\frac{r/r_S}{\dot{m}}\right)+6}{\left(\frac{r/r_S}{\dot{m}}\right)^2(1+\dot{m}^{-1}+6\dot{m}^{-2})}\,p_{ph}\,,
\end{equation}
where $p_{ph}$ is the value of the pressure at the photosphere, set by the boundary conditions.
Below the trapping radius, where $r/r_S \ll 3\dot{m}$, the third term in the numerator dominates and the pressure scales like $r^{-2}$, as expected from the adiabatic limit in free-fall. Outside the trapping radius, where $r/r_S \gg 3\dot{m}$, the pressure asymptotes to a constant so that the flow is isothermal at large radii. A comparison between an exact numerical integration of the non-linear equation and the analytic solution is shown in Fig \ref{fig:IsothermalSolutions} for model parameters of $\dot{m} = 700$, $M = 10 M_\odot$ and a photospheric temperature of $T = 10^4$K.

Higher order corrections can subsequently be computed from Eq \eqref{eq:ODESystemBLx1} using the solution for $\chi_0$ in Eq \eqref{eq:chi0_BLx1}. In principle, the boundary layer solution also needs to be supplemented by an `outer' solution that describes the flow outside the boundary layer, and is necessary to exactly satisfy the boundary condition at the inner boundary. However, the correction from the outer solution is small and only affects subleading orders, as Eq \eqref{eq:chi0_BLx1} by itself asymptotes to the required adiabatic limit at small $x$. Moreover, we are interested in the dynamics of the flow at large radii, to which the outer solution only has a negligible contribution. Since the leading order solution is already an excellent approximation to the exact one, as can be seen in Fig \ref{fig:IsothermalSolutions}, it is not necessary to include higher order corrections nor the outer solution. For completeness, the two leading orders of the outer solution are derived in Appendix \ref{App:OuterBLSol}.

Far outside the trapping radius, the rest-frame luminosity increases like $r^{1/2}$ and the diffusive luminosity $L_d = r^2\Delta L'(r)$ scales as $r^{3/2}$. The scaling $L_d\propto r^{3/2}$ is trivial and can be obtained from the entropy equation in the free-fall limit assuming a constant pressure. One immediately arrives at $\partial L_d/\partial m \sim const$ where $m$ is the Lagrangian mass coordinate, and therefore $L_d\propto r^{3/2}$ from mass conservation. How can we understand the scaling of the rest-frame luminosity? From Eq \eqref{eq:Bernoulli} we see that any change in the rest frame luminosity comes from the term $\left[v^2/2-GM/(r-r_S)\right]'$. Using the zeroth order result $L_d \propto r^{3/2}$ in Eq \eqref{eq:MomentumEqn}, we have $\Delta L'\propto\left[v^2/2-GM/(r-r_S)\right]'\propto r^{-1/2}$, and upon integration one obtains $\Delta L\propto r^{1/2}$. This is a consequence of the dynamical effect of the non-vanishing pressure on the flow, beyond the pressure-less free-fall limit.


The fact that the temperature does not increase with compression can be understood as an analogue of latent heat: since radiation enthalpy is the main source of energy and photons can escape freely outside the trapping radius, any increase in the temperature due to compression is followed by an immediate `cooling' via photons carrying a comparable energy flux. The gravitational field in this case is not a significant heating source, as noted by \cite{Vitello1984}, but merely facilitates the conversion of inward advected flux into diffusive flux through isothermal compression. As evident from Fig \ref{fig:IsothermalLum}, the internal energy per unit mass in the diffusive region decreases with compression, rather than increases.

The ratio between the rest-frame luminosity and the diffusive luminosity near the photosphere is
\begin{equation}\label{eq:delL_Ld_ph}
    \frac{\Delta L}{L_d}\bigg|_{ph} = \delta\left(1-6\delta\right) + \mathcal{O}(\delta^{3/2})\,.
\end{equation}
Eq \eqref{eq:delL_Ld_ph} shows that the luminosity that escapes the photosphere is a factor of $\dot{m}$ smaller than the co-moving diffusive luminosity there.
\begin{figure}
    \centering
    \includegraphics[width=0.99\linewidth]{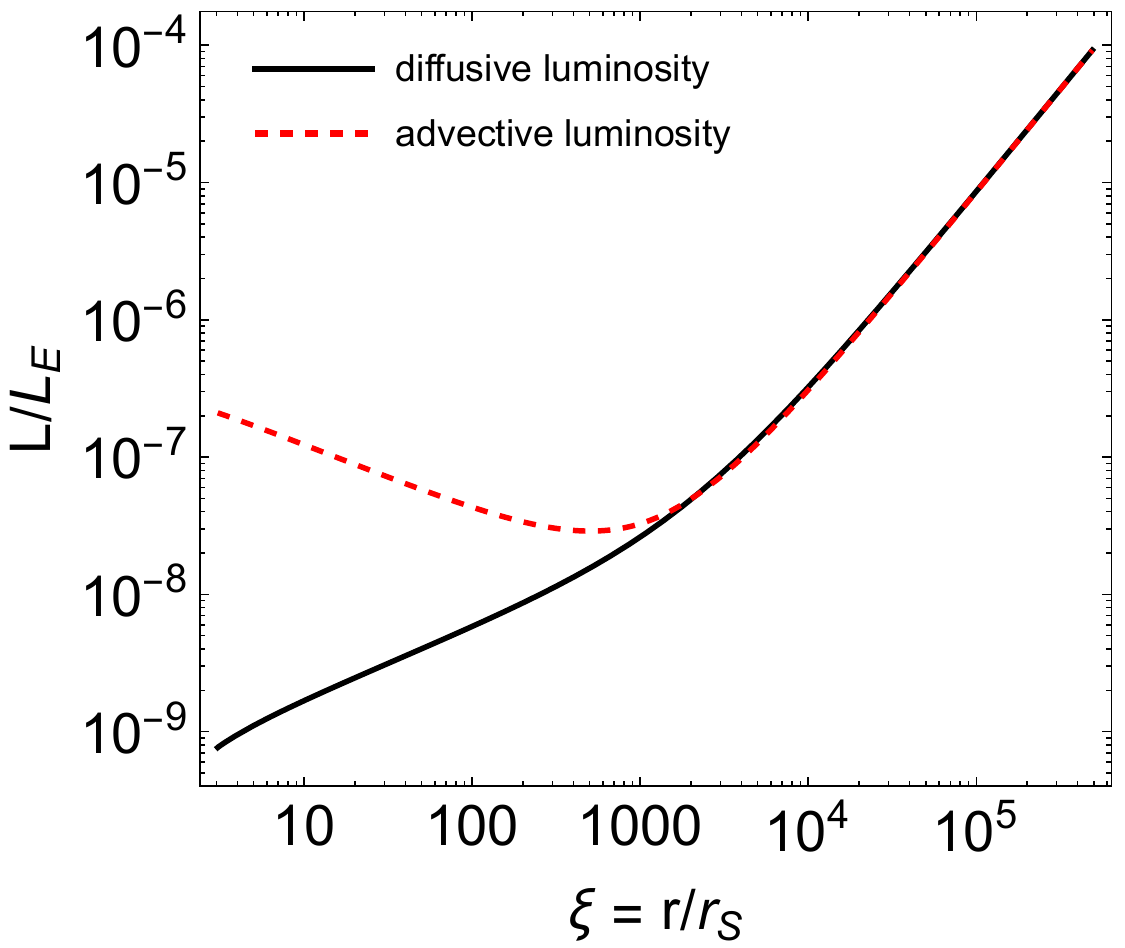}
    \caption{Comparison between the diffusive and the advective luminosities for the same conditions of Fig \ref{fig:IsothermalSolutions}. Beyond the trapping radius the two luminosities are comparable and the rest-frame luminosity, which is the difference between the diffusive and advective luminosity, is a factor of $\dot{m}^{-1}$ smaller.}
    \label{fig:IsothermalLum}
\end{figure}

\subsection{Powerlaw pressure solutions, $z_{bf}>0$}
In these solutions there is no boundary layer near $x=1$, since the boundary conditions dictate that $\chi$ asymptotes to a constant. All coefficients $C_n$ in \S \ref{sec:BLAnalysisIsothermal} are required to vanish and we must search for other solutions of Eq \eqref{eq:ChiODELinearized}. As we show in Appendix \ref{App:OuterBLSol}, all leading terms in the perturbative outer solution diverge near $x=1$, which means that all coefficients $B_n$ of the outer solution must vanish as well, in order to satisfy the boundary condition $\chi(x\rightarrow1) = const$. An accurate analytic solution is therefore difficult to obtain, and we aim at computing only the leading order solution for which $\chi$ is bounded at $x=1$. 

\begin{figure}
    \centering
    \includegraphics[width=0.98\linewidth]{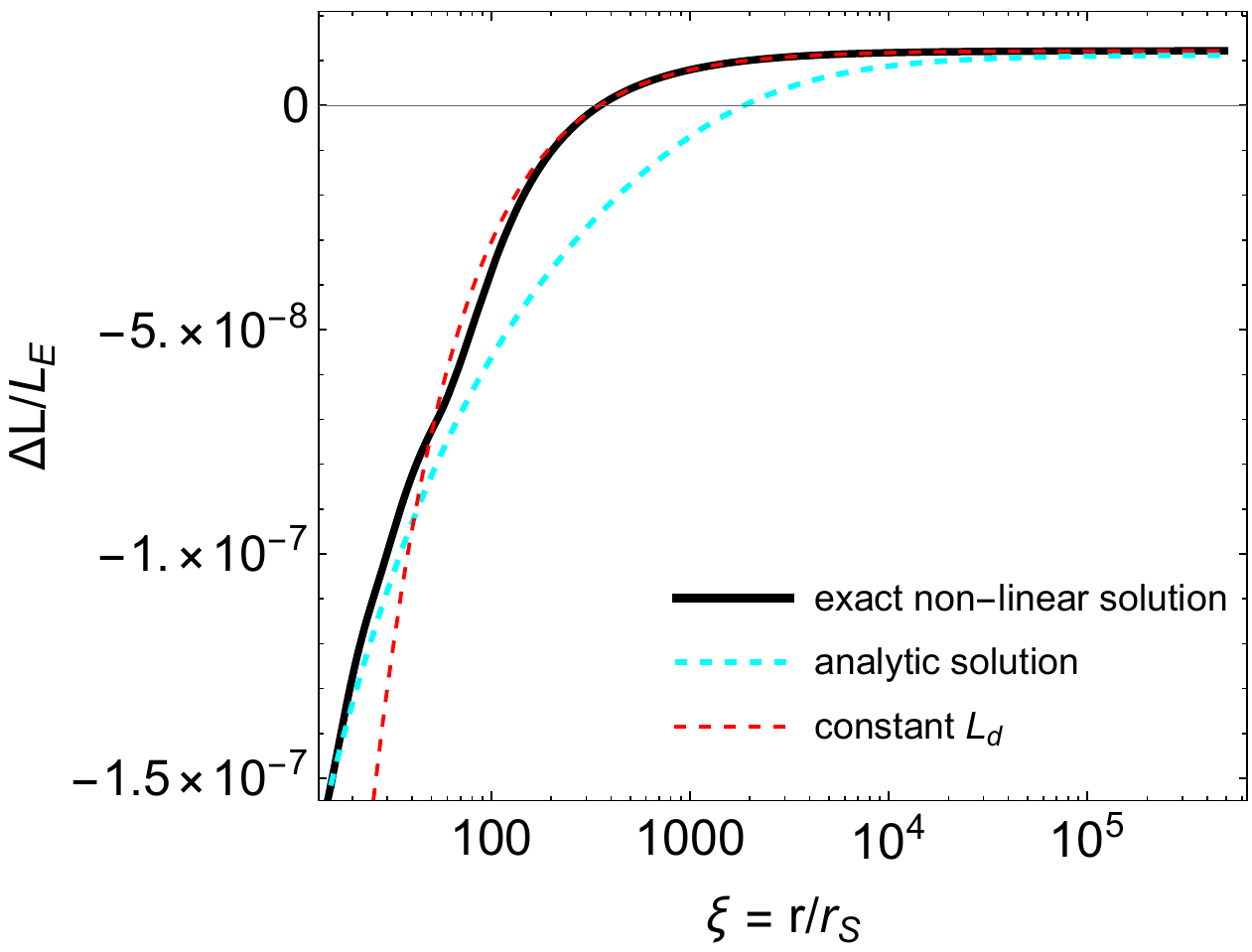}
    \includegraphics[width=0.98\linewidth]{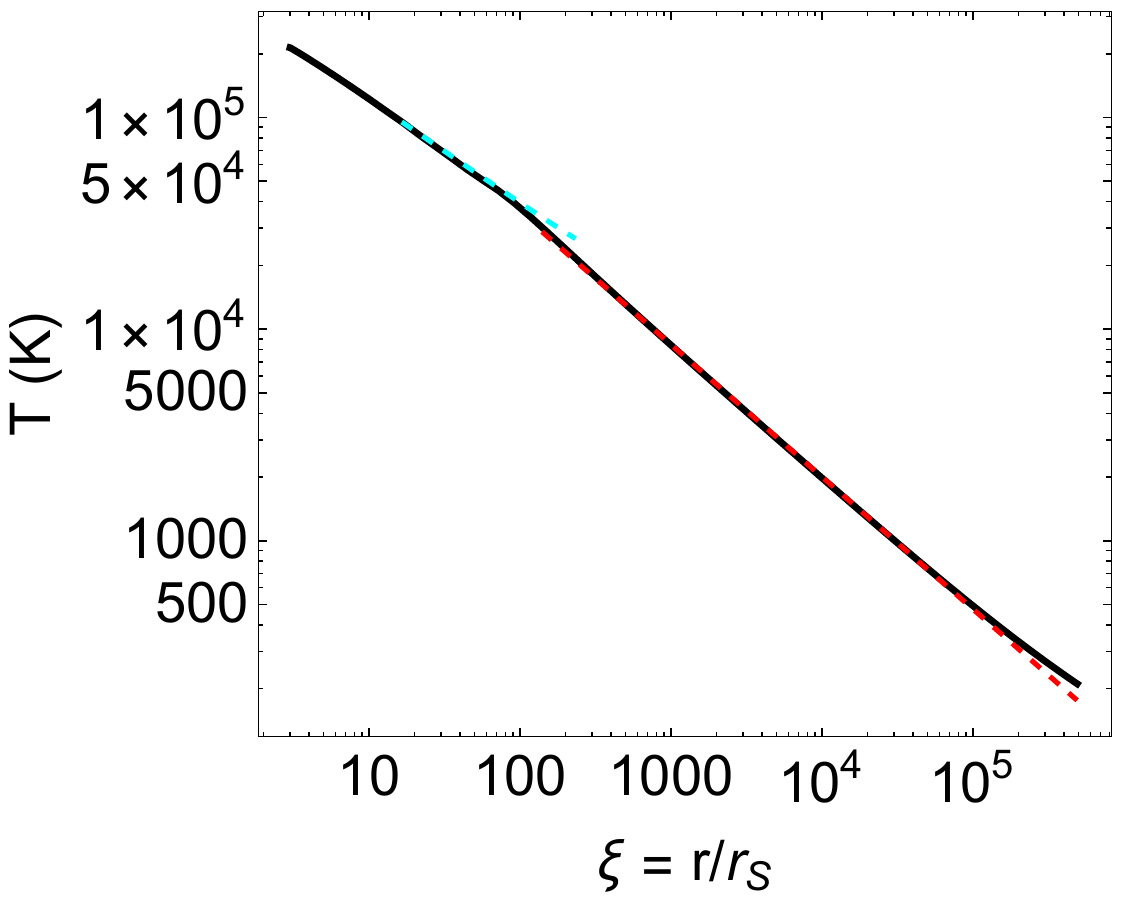}
    \caption{Leading order analytic solution (Eq \ref{eq:Chi0AnalyticInnerBL}) and exact non-linear numerical solution for the low temperature branch, using $\dot{m} = 700$ and $M = 10  M_\odot$. The analytic solution does not describe the temperature well beyond the trapping radius, as higher order corrections are required, but provides a decent fit for the rest-frame luminosity.}
    \label{fig:PowerLawSolutions}
\end{figure}

As explained in the discussion at the end of \S \ref{sec:BifurcatingSolutions}, an internal boundary layer, located away from the boundaries of the domain, exists if the coefficient of $\chi'$ in Eq \eqref{eq:ChiODELinearized}, $a(x)$, vanishes somewhere in $0<x<1$. This indeed occurs at $x = x_{bf}$ (see Eq \ref{eq:x_bf}). Let us investigate the immediate vicinity of $x_{bf}$. For that, we will follow the analysis in section 9.6 of \cite{BenderOrszag}. The boundary layer is characterized by a thickness $\Delta\ll1$, which can be greater or smaller than $\delta$. From a dominant balance analysis, we again find that $\delta \sim\Delta$ up to a logarithmic order correction. We make a transformation to a boundary layer variable
\begin{equation}\label{eq:defXBLx1}
    X \equiv \frac{x-x_{bf}}{\delta}\,~; ~ \chi(x) = Y(X)\,.
\end{equation}
Here $X=1$ at the outer boundary and $X\rightarrow-\infty$ at the inner boundary.
To leading order around $X=0$, we can approximate Eq \eqref{eq:ChiODELinearized} as
\begin{equation}
    Y_0''+\alpha X~Y'_0+\beta ~Y_0 = 0\,,
\end{equation}
with $\alpha = -3/2$ and $\beta = 3/4$.
Now, setting $Y_0 = e^{-\alpha X^2/4}W$ and $\sqrt{\alpha}X = Z$, we obtain the parabolic cylinder equation:
\begin{equation}\label{eq:ParabolicCylinderEqn}
    W''+\left(\frac{\beta}{\alpha}-\frac{1}{2}-\frac{1}{4}Z^2\right)W = 0\,.
\end{equation}
The solution to Eq \eqref{eq:ParabolicCylinderEqn} is a linear combination of the parabolic cylinder functions:
\begin{multline}
    Y_0(X) = e^{-\alpha X^2/4}\big\{C_1 D_{-\beta/\alpha}(X\sqrt{-\alpha}) \\ +C_2 D_{-\beta/\alpha}(-X\sqrt{-\alpha})\big\}\,,
\end{multline}
where $C_1, C_2$ are constants. The parabolic cylinder functions obey the following asymptotic behaviour:
\begin{equation}
\begin{aligned}
    &D_\mu(t) \propto t^{\mu} e^{-t^2/4} &\,,~t\rightarrow+\infty\\
    &D_\mu(-t) \propto t^{-\mu-1} e^{t^2/4} \frac{\sqrt{2\pi}}{\Gamma(-\mu)}~&\,,~t\rightarrow+\infty
\end{aligned}~~\,.
\end{equation}
Here, $\mu = -\beta/\alpha = 1/2$. We can immediately see that a problem arises due to the fact that $\alpha = -3/2<0$: $X$ varies from $-\infty$ to $1$ and $e^{-\alpha X^2/4}$ increases exponentially for large $X$. Now, since $D_{-\beta/\alpha}(X\sqrt{-\alpha})$ diverges as $X\rightarrow-\infty$, we must set $C_1 = 0$. We are left with
\begin{equation}
     Y_0(X) = C_2\, e^{-\alpha X^2/4} D_{-\beta/\alpha}(-X\sqrt{-\alpha})\,.
\end{equation}
The leading order solution agrees with the general requirements on the boundary conditions, namely $Y_0'(X\rightarrow-\infty) = 0$, $Y_0'(X \rightarrow1) = const$ and $Y_0(X\sim1)$ being finite. Moreover, as $X\rightarrow-\infty$, $\chi \propto (x-x_{bf})^{1/2}$ and $\chi'\propto (x-x_{bf})^{-1/2} \propto \left(\frac{\xi}{1+\xi/\dot{m}}\right)^{1/2}$. This scaling agrees with the expectation in the adiabatic limit, where $\xi \ll \dot{m}$.
Changing back to the original variable $x$ via $X = \frac{x-x_{bf}}{\delta}$, the leading order solution for $\chi$ is
\begin{equation}\label{eq:Chi0AnalyticInnerBL}
    \chi_0 =C_2\, D_{1/2}\left[\sqrt{\frac{3}{2}}\frac{x-x_{bl}}{\delta}\right]e^{\frac{3}{8}\left(\frac{x-x_{bl}}{\delta}\right)^2} \,.
\end{equation}
The rest frame luminosity terms of $r$ is
\begin{multline}
    \Delta L(r) = -\frac{D_{1/2}\left[-\sqrt{\frac{3}{2}}\left(1-\frac{\dot{m}}{r/r_S}\right)\right]}{2D_{1/2}\left[-\sqrt{\frac{3}{2}}\right]}\\ \times e^{\frac{3}{8}\left(\frac{r/r_S}{\dot{m}}\right)^{-2} \left(1-\frac{2\,r/r_S}{\dot{m}}\right)}\Delta L_\infty\,,
\end{multline}
with $\Delta L_\infty$ being the rest-frame luminosity at infinity. Here, the rest-frame luminosity is effectively equal to the diffusive luminosity, since the pressure vanishes at infinity.
In Fig \ref{fig:PowerLawSolutions} we compare the exact numerical solution of Eq \eqref{eq:PsiODE} with the leading order solution of Eq \eqref{eq:Chi0AnalyticInnerBL}. Although $\chi_0$ reproduces the general limiting behaviour of the solution, it is not as good of an approximation as that obtained for the isothermal solutions, and higher orders of $\chi_n$ must be computed to reproduce all of the characteristics of the solution. In the same figure we overplot the asymptotic solution of Eq \eqref{eq:ChiConstLd}, representing a constant diffusive luminosity. It appears that the exact solution is a non-trivial combination of the two solutions, and that Eq \eqref{eq:ChiConstLd} accurately describes the flow outside the trapping radius, while Eq \eqref{eq:Chi0AnalyticInnerBL} provides a good fit inside the trapping radius. As the diffusive luminosity approaches a constant, the pressure acquires the limiting scaling of $p\propto r^{-5/2}$. Near the outer boundary, the scale height to change $\Psi$ is much greater than the local radius, an immediate inference from the fact that $\chi = -2\Delta \Lambda$ is asymptotically constant.

Contrary to the isothermal branch of the solutions, here the pressure does not play an important role in the dynamics at large radii, and the correction to rest-frame luminosity due to deviation from the free-fall limit is negligible.

\begin{figure}
    \centering
    \includegraphics[width=0.99\linewidth]{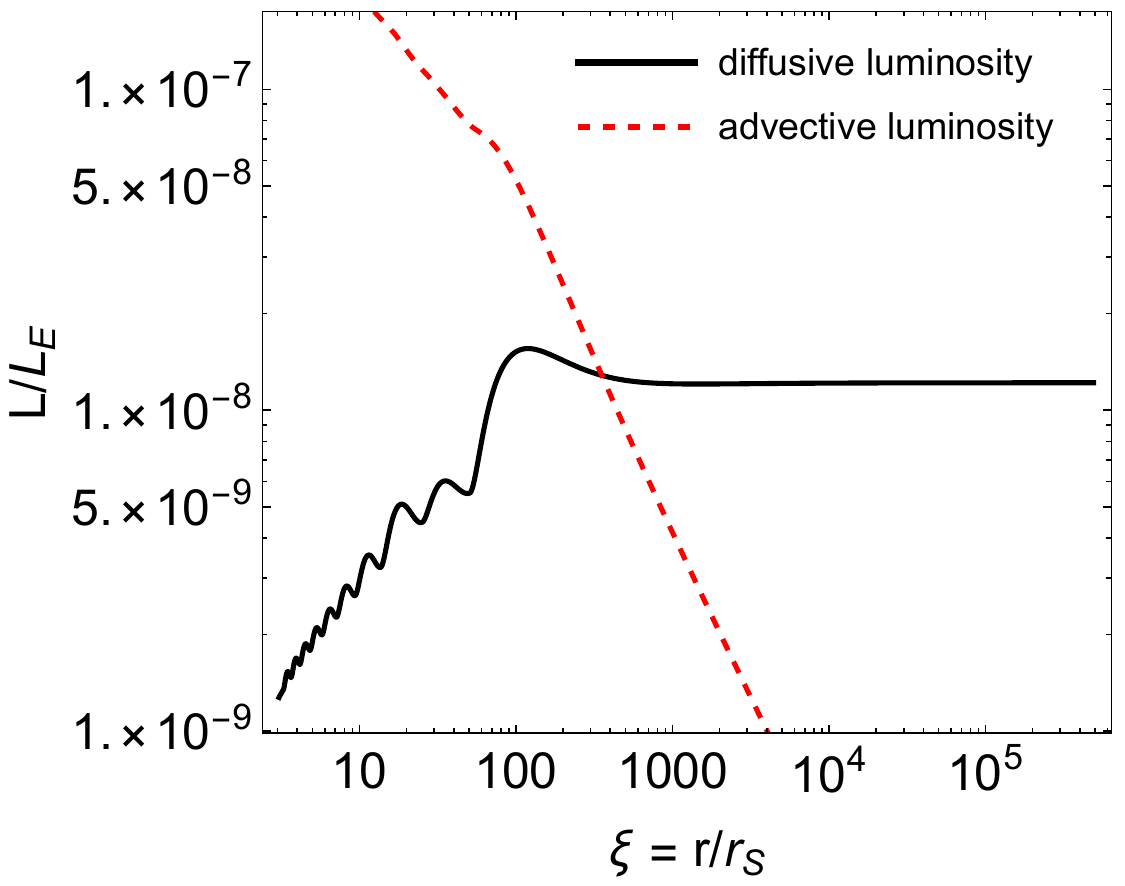}
    \caption{Comparison between the diffusive and the advective luminosities derived from exact numerical integration of Eq \eqref{eq:PsiODE} for the same conditions of Fig \ref{fig:PowerLawSolutions}. Beyond the trapping radius the rest-frame luminosity is equal to the diffusive luminosity.}
    \label{fig:PowerLawLum}
\end{figure}

\section{Fallback accretion following weak stellar explosions}\label{sec:Applications}
We now apply the results of the previous sections to study fallback accretion flows following failed supernova events. The gravitational collapse of a stellar core is typically followed by the formation of a strong shock wave that disrupts and unbinds the entire stellar envelope, carrying $\sim 10^{51}$ erg. 
%
However, as discussed in the introduction, not all massive stars necessarily produce an energetic explosion following core-collapse.  Even state-of-the-art 3D simulations find that some progenitors fail to explode \citep{Burrows2024}.  The results of current explosion simulations also depend sensitively on structure of the progenitor star, which remains uncertain.    

Core-collapse that does not produce an energetic neutrino-powered explosion may still produce a unique electromagnetic signal: it has been suggested by \cite{Nadezhin1980} (and subsequently studied by e.g., \citealt{Lovegrove2013,Lovegrove2017,Fernandez2018,Ivanov2021}), that the hydrodynamic response of the stellar envelope due to the loss of $0.2-0.5$ M$_\odot$ to neutrinos during the proto-neutron star phase leads to an acoustic perturbation in the outer envelope of the star, which eventually steepens into a weak shock wave. The shock can accelerate as it propagates down the stellar envelope if it encounters a steep density gradient, such as the edge of the the hydrogen envelope or the edge of the helium core. In this case, the shell that is marginally bound to the black hole after shock passage is located where the Mach number of the shock becomes of order unity, at the transition to a strong shock.  An alternative mechanism for producing a low energy explosion during otherwise `failed' core-collapse is that net rotation of the stellar envelope and/or random angular momentum from large turbulent motions in a convective supergiant lead to accretion and outflows that power an explosion (e.g., \citealt{Gilkis2014,Antoni2023}).  

In order to model fallback accretion, we assume a `weak' stellar explosion in which part of the material traversed by the outgoing shock wave becomes unbound and escapes the gravitational field of the star, while inner mass shells remain highly bound and instantaneously collapse to form a black hole. Shells that are less bound after shock passage move outwards, reach a maximum height where $v=0$ (hereafter referred to as the turnaround radius), and eventually fall into the newly formed black hole. At asymptotically late times, the entire space between the turn-around radius and the black hole is occupied by a single mass shell that is marginally bound to the black hole. We shall assume that at the initial characteristic time $t_0$ a shell located at $r_0$ is expanding with velocity $v_0 = r_0/t_0$. This time-dependent accretion flow becomes steady at small radii, and eventually acquires an accretion rate $\dot{M}(t)$ that is independent of the radius. If we assume that to leading order, the pressure plays a negligible role in the dynamics at very large radii outside the photosphere, mass elements follow ballistic trajectories and one can solve for the asymptotic accretion rate of the steady flow \citep{Chevalier1989}:
\begin{equation}\label{eq:MdotChevalier}
    \dot{M} = \frac{8}{9}\pi^{5/3}(2 G M)\rho_0 t_0^{8/3} t^{-5/3}\,,
\end{equation}
where $\rho_0$ is the initial density of the shell that is marginally bound to the black hole.
Using \cite{Chevalier1989}'s equation 3.20 for the density in the interior region of the flow
\begin{equation}\label{eq:RhoInnerChevalier}
    \rho_1 = \frac{2}{9}\pi^{2/3}(2G M)^{1/2}\rho_0 t_0^{8/3} t^{-5/3} r^{-3/2}\,,
\end{equation}
we confirm that the assumption $p_r \gg p_g$ near the photosphere is indeed satisfied:
\begin{equation}\label{eq:PrPg_mdot}
    \frac{p_r}{p_g}\bigg|_{ph} \simeq 7200 \left(\frac{\dot{m}}{10^4}\right)^{2}\left(\frac{M}{10 M_\odot}\right)\left(\frac{T_{ph}}{10^4 K}\right)^{3}\,.
\end{equation}

To determine which of the solution branches discussed in this paper describes fallback accretion, we need to estimate the scale height of $\Psi$ near the photosphere. We will work under the assumption that at large radii the pressure plays a sub-dominant role in the momentum equation, and use \cite{Chevalier1989}'s results to solve for the temperature above which the change in $\Psi$ is of order $\Psi$. Using \cite{Chevalier1989}'s equation 3.8, we find that $\epsilon \simeq r/r_1$, where $r_1$ is the maximal extent reached by a fluid element, and is a function of $r_0, t_0$ and $M$. Since the time to reach maximum height is approximately equal to the time it takes a marginally bound fluid element to fall back down to the black hole, we can approximate $r_1$ as the turnaround radius at time $t$, so that
\begin{equation}
    \Psi_{p=0}\simeq \frac{r_S}{r_{ta}} = \frac{r_S}{(8 G M t^2/\pi^2)^{1/3}}\,.
\end{equation}
Comparing the above estimate for $\Psi$ to $4p/(\rho c^2)$, we find that the solution tends to become isothermal if
\begin{multline}
    T_{ph}\gtrsim 2100 \text{K} \left(\frac{M}{10 M_\odot}\right)^{1/20}\left(\frac{r_0}{10 R_\odot}\right)^{-2/5}\\\left(\frac{\rho_0}{10 ^{-5}\frac{g}{cm^3}}\right)^{1/10}\times\left(\frac{\dot{m}}{10^4}\right)^{-2/5}\,.
\end{multline}
The solution will thus likely settle onto the isothermal branch, as long as the requirement for a fully ionized flow is satisfied.

In our simplified model that neglects radiative heating processes, the only source of heating in the inflowing material is compressional work done on the gas by the gravitational field. The heating rate due to compression outside the photosphere is $\Gamma_H = 8\times 10^{-26}\, T_4 \,\dot{m}^{-1} \text{erg} \, \text{cm}^3 \, \text{s}^{-1}$, taking into account the fact that gas and radiation are decoupled there. In the optically thin region, 
the cooling rate for a hydrogen dominated plasma drops steeply at a temperature of $10^4$K \citep[][]{Buff1974,Schure2009} from a level of $\sim 10^{-21} \text{erg} \, \text{cm}^3 \, \text{s}^{-1}$. Equilibrium between heating and rapid cooling processes therefore dictates that the photospheric temperature lies a bit below $10^4$K, which was also found by \citet{Blondin1986}.
For very high values of $\dot{m}$ the heating rate is a few orders of magnitude lower than the lowest value in the available cooling curves.  It is therefore likely that the photospheric temperature is even lower than $\sim 10^4$ K.   In this case the gas is only partially ionized and our assumption of electron scattering opacity breaks down (we return to this below). The photosphere of a fully ionized gas will enter the accretion region once the accretion rate falls below 
\begin{equation}\label{eq:mdot_thin_ta}
   \dot{m}_{\tau =1, ta} \simeq 10^4 \left(\frac{M}{5 M_\odot}\right)^{-1/2}\left(\frac{r_0}{10 R_\odot}\right)^{2/3}\left(\frac{\rho_0}{10^{-5} g \, cm^{-3}}\right)^{1/6}\,.
\end{equation}
This accretion rate represents a lower limit for the onset of recombination in the inflowing material, which corresponds to an upper limit on the time of recombination:
\begin{equation}\label{eq:trec}
\begin{split}
    t_{rec} &\leq 260 \text{d} \,\left(\frac{M}{5 M_\odot}\right)^{-1/2}\left(\frac{r_0}{10 R_\odot}\right)^{2}\left(\frac{\rho_0}{10^{-5} \text{g cm}^{-3}}\right)^{1/2} \,.
\end{split}
\end{equation}
We discuss the case of low photospheric temperatures that lead to recombination in \S \ref{sec:recombination}.

\subsection{The luminosity from a fully ionized inflow}
Assuming photospheric temperatures greater than $\sim 10^4$ K to satisfy a fully ionized flow, super-Eddington fallback accretion flows tend to be radiation pressure dominated near the photosphere at early times when $\dot m$ is large. As the accretion rate decreases with time, the point where gas and radiation pressure are comparable moves outwards. The characteristics of the observed radiation will change substantially once gas pressure becomes dominant near the photosphere, and the luminosity at later times will be dominated by gas thermal energy that diffuses out of the trapping radius. We compute the observed signal in the two regimes.

A naive estimate for the observed luminosity is the rest-frame luminosity at the photosphere:
\begin{multline}
   \Delta L_{ph}/L_E = 4 \Pi_{ph} \, \dot{m}^3 \\ \simeq 3 \times 10^{-5} \left(\frac{\dot{m}}{10^4}\right)^2\left(\frac{T_{ph}}{10^4 \text{K}}\right)^4 \left(\frac{M}{10 M_\odot}\right)\,  \propto t^{-10/3}\,.
\end{multline}
However, this neglects the emission from the optically thin region exterior to the photosphere:   the luminosity can keep increasing in the optically thin region as a result of cooling of gas internal energy. Assuming that the gas temperature does not change outside the photosphere, as gas and radiation are decoupled, we can approximate the cooling luminosity by $L_c = \frac{5}{2}\frac{k_B T_{ph} }{m_p}\dot{M}$, or
\begin{equation}\label{eq:Lambda_c_gas}
    L_c/L_E \simeq 2 \times 10^{-9} \left(\frac{T_{ph}}{10^4 \text{ K} }\right)\dot{m} \propto t^{-5/3}\,.
\end{equation}
Comparing the two contributions, the cooling luminosity in the optically thin region dominates when
\begin{equation}\label{eq:mdot_c}
    \dot{m} \lesssim \dot{m}_c =  8700\, \left(\frac{T_{ph}}{10^4 \text{K}}\right)^{-3}\left(\frac{M}{10 M_\odot}\right)^{-1}\,.
\end{equation}
If the accretion rate is initially greater than $\dot{m}_c$, the luminosity first decreases rapidly as $t^{-10/3}$ and flattens to $t^{-5/3}$ once $\dot{m} = \dot{m}_c$. At even later times, $p_r < p_g$ in the diffusive region, and the luminosity is governed by gas internal energy that diffuses out of the trapping radius, located approximately at
\begin{equation}\label{eq:rtr_pg}
    r_{tr} \sim \frac{p_g}{4p_r}\frac{\kappa}{c}\dot{M}\,.
\end{equation}
By definition, the flow just inside the trapping radius is adiabatic, and therefore $T\propto\rho^{2/3}$. Using Eq \eqref{eq:RhoInnerChevalier}, we find that $r_{tr}$ is independent of time once the flow is gas pressure dominated, $r_{tr} \sim \dot{m}_{eq} \times r_S$, where $\dot{m}_{eq}$ is the value of $\dot{m}$ at the moment when $p_g=p_r$ near the trapping radius. The observed luminosity is approximately $L_{obs}\sim \frac{5}{2}\frac{k_B T}{\mu}\dot{M}\propto \dot{m}^{5/3}\propto t^{-25/9}$. The three regimes are shown in Fig \ref{fig:L_fallback}.

\subsection{The effect of recombination and ionization on the luminosity of fallback accretion}\label{sec:recombination}
As the marginally bound mass shells expand out towards the turnaround radius, the flow is initially opaque and fully ionized, and its temperature falls adiabiatically. The evolution of the temperature will deviate from adiabaticity once the point where $\tau = c/v$ crosses the marginally bound shell (at $\sim 1.77 r_{ta}$) and the inflow becomes diffusive ($\dot{m} = \dot{m}_{tr}$).
The temperature will, eventually, drop below the recombination temperature of hydrogen; as a result, the entire isothermal region of the flow will recombine rapidly and release the internal energy of the material above the trapping radius, resulting in a luminous flare. A weak constraint on the onset of recombination comes from when the electron scattering photosphere crosses the turnaround radius ($\dot{m} = \dot{m}_{\tau=1,ta}$), as ionization cannot be sustained by compressional work near the photosphere at high accretion rates.

The dynamics of the recombination front cannot be described as part of the steady-state flow, since it traverses the material on a time scale much shorter than the dynamical time of the system. However, once the Eulerian velocity at the recombination front becomes comparable to the local fluid velocity, the recombination front changes its role to an ionization front, as it is crossed by neutral gas that has previously recombined. The ionization front is now part of the steady flow, and its dynamics are dictated by the balance of the escaping photon energy flux with the energy flux required to ionize the incoming material, namely, $a c T_{rec}^4/\tau_{rec}  \sim v\rho \chi_i/m_p$, where $\chi_i$ is the ionization potential and $\tau_{rec}$ is the optical depth where $T=T_{rec}$. We now assume that the gas and radiation decouple as soon as the gas temperature drops below $T_{rec}$; the gas temperature keeps decreasing with the ionization fraction while the radiation temperature remains fixed around $T_{rec}$. Following \cite{Faran2019}'s description of the recombination layer, we approximate $\tau_{rec} \sim (T_{rec}/T_{g,ph})^4$, where $T_{g,ph}$ is the gas temperature at the photosphere. The observed luminosity is then $L\sim 4\pi r_{ion,r}^2 a c T_{rec}^4$:
\begin{equation}\label{eq:L_ion_r}
    L_{ion,r}/L_E \sim \frac{\chi_i}{m_p c^2} \tau_{rec} \dot{m} \simeq 1.5 \times 10^{-8} \left(\frac{T_{rec}}{T_{g,ph}}\right)^{4}\dot{m}\,,
\end{equation}
where the photospheric gas temperature is typically of order $\sim T_{rec}/2$ (see e.g., \citealt{Zampieri1998}).


The material that crosses the ionization front becomes dominated by gas pressure once $\rho k_B T_{rec}/m_p \sim a T_{rec}^4$, at an accretion rate of $\dot{m} \simeq 43 M_{10}(T/T_{rec})^{12}$, where $T\sim T_{rec}$. The strong dependence on the temperature and the fact that $T_{ph}\sim T_{rec}/2$ is only an estimate leads to uncertainty in the exact transition value of $\dot{m}$.

\begin{figure}
    \centering
    \includegraphics[width=1\linewidth]{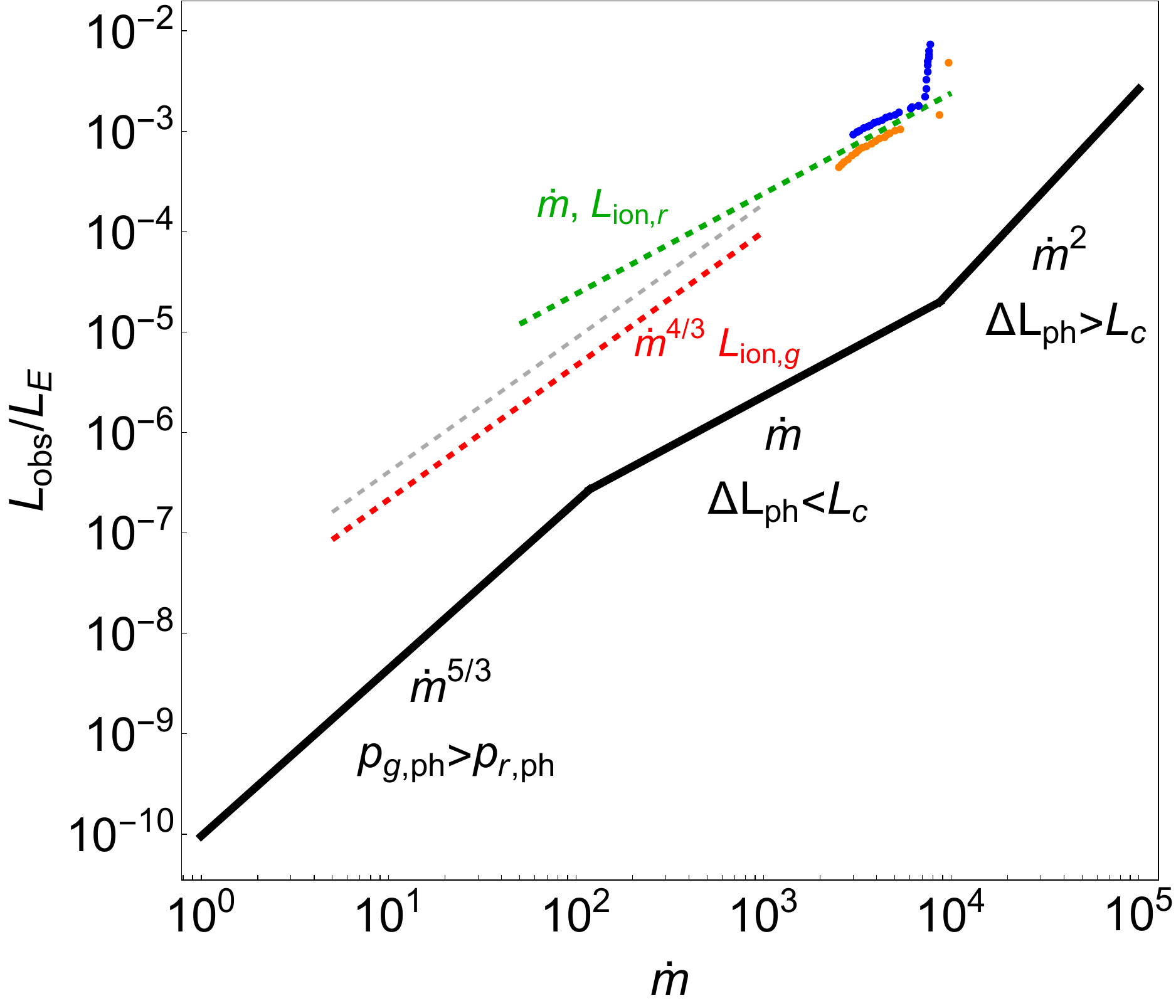}
    \caption{The observed luminosity from fallback accretion onto a $M = 10 M_\odot$ black hole for a photospheric temperature $T_{ph} = 10^4$ K, as a function of accretion rate $\dot m$ relative to Eddington (defined in eq. \ref{eq:mdotEdd}). The black line is the prediction for the luminosity emitted by a flow that is fully ionized near the photosphere, while the dashed lines are the predictions for a neutral inflow that goes through ionization in the limits $p_r \gg p_g$ (green; Eq \ref{eq:L_ion_r}) and $p_g \gg p_r$ (red; Eq \ref{eq:L_ion_g}). Numerical simulation results for accretion onto a $1.5 M_\odot$ black hole from \cite{Zampieri1998} are shown as blue and orange data points for their two models with recombining/ionizing envelopes. Eq \eqref{eq:L_ion_g} is also plotted in dash-gray for $M=1.5 M_\odot$ to compare with \cite{Zampieri1998}'s results.}
    \label{fig:L_fallback}
\end{figure}

A gas pressure dominated flow that goes through ionization is characterized by an effective adiabatic index of $\gamma_3 \simeq 1.1$ in the limit $p_g \gg p_r$. As a result, an extended, nearly isothermal region is formed, throughout which the ionization fraction decreases very slowly with the radius. The ratio $p_r/p_g$ increases with the radius in the isothermal region as the density decreases. Once $p_g\sim p_r$, a further decrease in density leads to a significant drop in the temperature, and the ionization fraction drops rapidly to $0$. The point where $p_r\sim p_g$ defines the effective photospheric radius, which follows the scaling $r_{ion,g}\propto T^{-2} \dot{m}^{2/3}$.
The escaping luminosity in this case will be
\begin{equation}\label{eq:L_ion_g}
    L_{ion,g}/L_E = 4\pi\, r_{ion,g}^2 a c\, T_{ion,g}^4 \simeq 10^{-8} \dot{m}^{4/3}M_{10}^{-1/3}\,.
\end{equation}
Our results are in agreement with \cite{Zampieri1998}'s numerical results for models of the radiation from fallback accretion with recombining envelopes, shown in Fig \ref{fig:L_fallback}. Even though both Eq \eqref{eq:L_ion_r} and \eqref{eq:L_ion_g} agree well with the simulations, the temperature profiles shown in \cite{Zampieri1998}'s Fig 7a do not show any extended ionization region, which indicates that the flow is radiation pressure dominated and their results therefore most likely follow Eq \eqref{eq:L_ion_r}.

We note that a similar scaling to Eq \eqref{eq:L_ion_g} was obtained by \cite{Blondin1986} despite not explicitly accounting for the effects of ionization. The similarity of the two results is because \cite{Blondin1986} forced $p_r = 4p_g$ near the trapping radius. 
An inspection of their analysis shows that this requirement relies on an inaccurate expression for the trapping radius, which, in general, should include a factor of $1+p_g/(4p_r)$ to account for the competing contributions of radiation and gas pressure. Not including the gas pressure term naturally (but incorrectly) forces $p_g\sim4p_r$ in the regime where $p_g\gtrsim p_r$.  

\subsection{{Application to Supergiants and Failed Supernovae Candidates}}
\label{sec:candidates}

\begin{figure}
    \centering
    \includegraphics[width=0.95\linewidth]{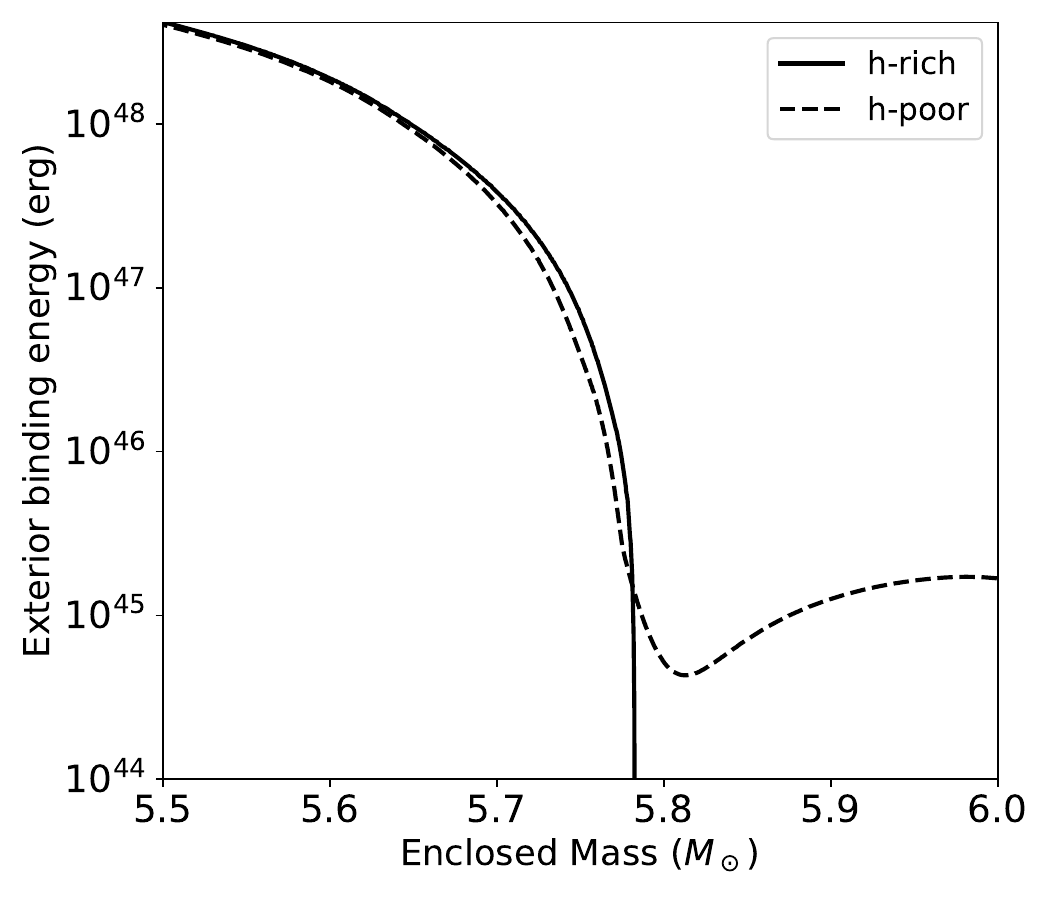}
    \caption{The binding energy of the material exterior to a given mass coordinate, including both the gravitational potential energy and the internal energy of the material, for our H-rich red supergiant model and H-poor yellow supergiant model. For a wide range of explosion energies, the mass of the black hole is $\simeq 5.5-5.8 M_\odot$ for this progenitor and the ejecta mass is the mass of the hydrogen envelope.}
    \label{fig:EbMESA}
\end{figure}

We now apply our results to the properties of a 14 $M_\odot$ (ZAMS) red supergiant (RSG) progenitor evolved to near core-collapse using MESA; we also consider a yellow supergiant (YSG) model obtained from the RSG model by stripping much of its hydrogen envelope (the models were kindly provided by Andrea Antoni from Antoni et al. in prep).  The RSG model has a mass of $13.2 M_\odot$ near collapse, a radius of $1090 R_\odot$, a He core mass of $5.6 M_\odot$, and a H envelope mass of $7.6 M_\odot$.   The YSG model has a mass of $6.1 M_\odot$ near collapse, a radius of $560 R_\odot$, a He core mass of $5.6 M_\odot$, and a H envelope mass of $0.5 M_\odot$.
In Fig \ref{fig:EbMESA} we plot the binding energy of the material exterior to mass $m(r)$, $E_b(>m) = -\int_{m(r)}^M \left[-G m'/r + u\right]dm'$, where $u$ is the specific energy per unit mass.
The shell that will be marginally bound after shock passage is traversed by the shock as it acquires a Mach number of $\sim$ a few, and has a binding energy comparable to the energy carried by the shock. The shock is presumed to originate from an acoustic perturbation that has steepened into a weak shock wave. This is a good approximation for the low energy explosions triggered by neutrino mass-loss considered in \citet{Lovegrove2013,Fernandez2018} (these explosions are spherically symmetric, consistent with our assumptions in this paper).  For low energy explosions driven by accretion of the convective hydrogen envelope, a better approximation is that the explosion always originates at the base of the hydrogen envelope \citep{Antoni2023}.   Calculating the fallback accretion rate in  this case is more complicated, as we discuss at the end of \S \ref{sec:candidates}. As weak shock waves evolve effectively adiabatically, the energy deposited in the unbound material is comparable to the initial energy of the acoustic perturbation, which is assumed to be $\lesssim 10^{49}$ erg.
The fact that the binding energy in the transition between the helium core and the hydrogen envelope is a very strong function of the enclosed mass implies that while the explosion energy can vary by orders of magnitude between $10^{45}-10^{48}$ erg, the mass that remains bound and forms a black hole changes very little and is always $\sim 5.6-5.8\, M_\odot$ for the model considered here.

\begin{figure}
    \centering
    \includegraphics[width=1\linewidth]{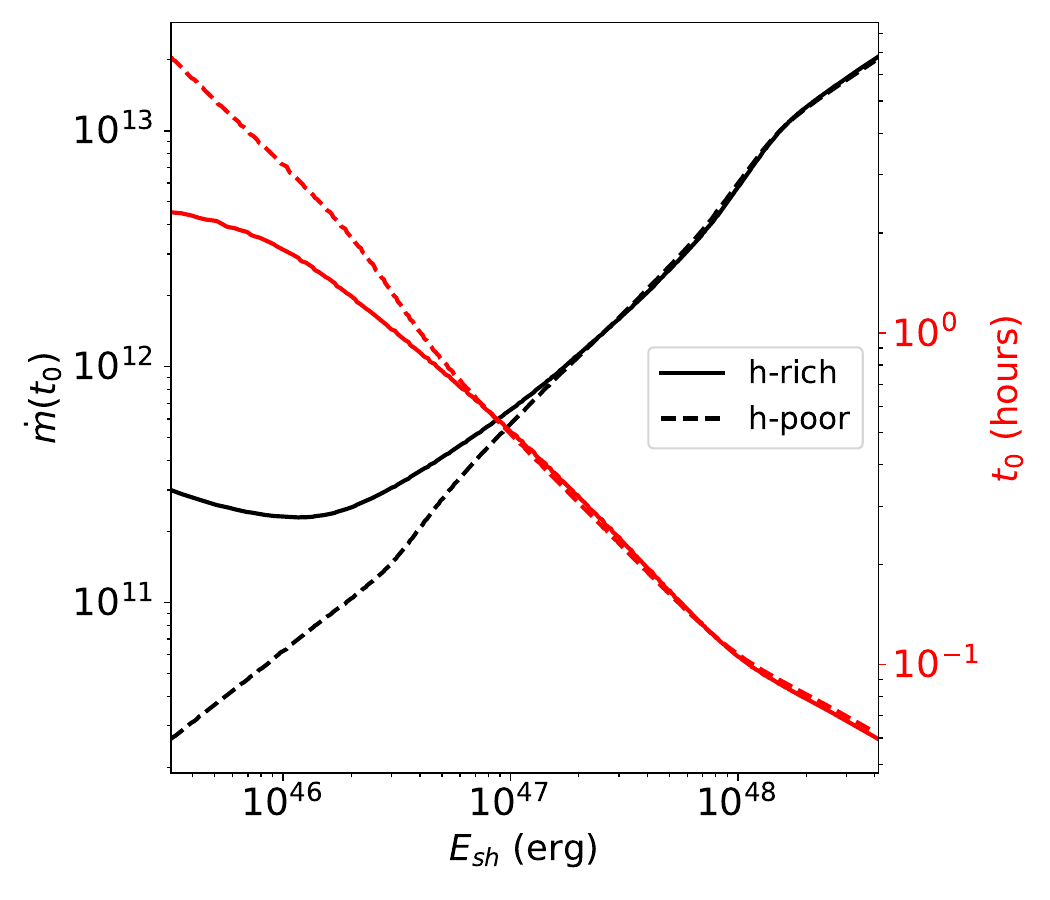}
    \caption{Estimated values for the initial accretion rate (in units of Eddington; Eq. \ref{eq:mdotEdd}) as a function of the explosion energy for our H-rich (red supergiant) and H-poor (yellow supergiant) MESA progenitor models (black). The corresponding initial free-fall time $t_0 = r_0/\sqrt{2GM/r_0}$ is shown in red.  The accretion rate decreases at late times with $\dot m \simeq \dot m_0(t/t_0)^{-5/3}.$}
    \label{fig:mdot_Eb_MESA}
\end{figure}

Using the density profiles of the progenitors and the assumption that the shock energy is comparable to the binding energy of the marginally bound shell, we use Eq \eqref{eq:MdotChevalier} to compute the initial accretion rates $\dot m(t_0)$ starting at time $t_0$ shown in Fig \ref{fig:mdot_Eb_MESA}; recall that for $t \gtrsim t_0$, $\dot m \simeq \dot m_0 (t/t_0)^{-5/3}$.

\begin{figure*}
    \centering
    \includegraphics[width=0.95\linewidth]{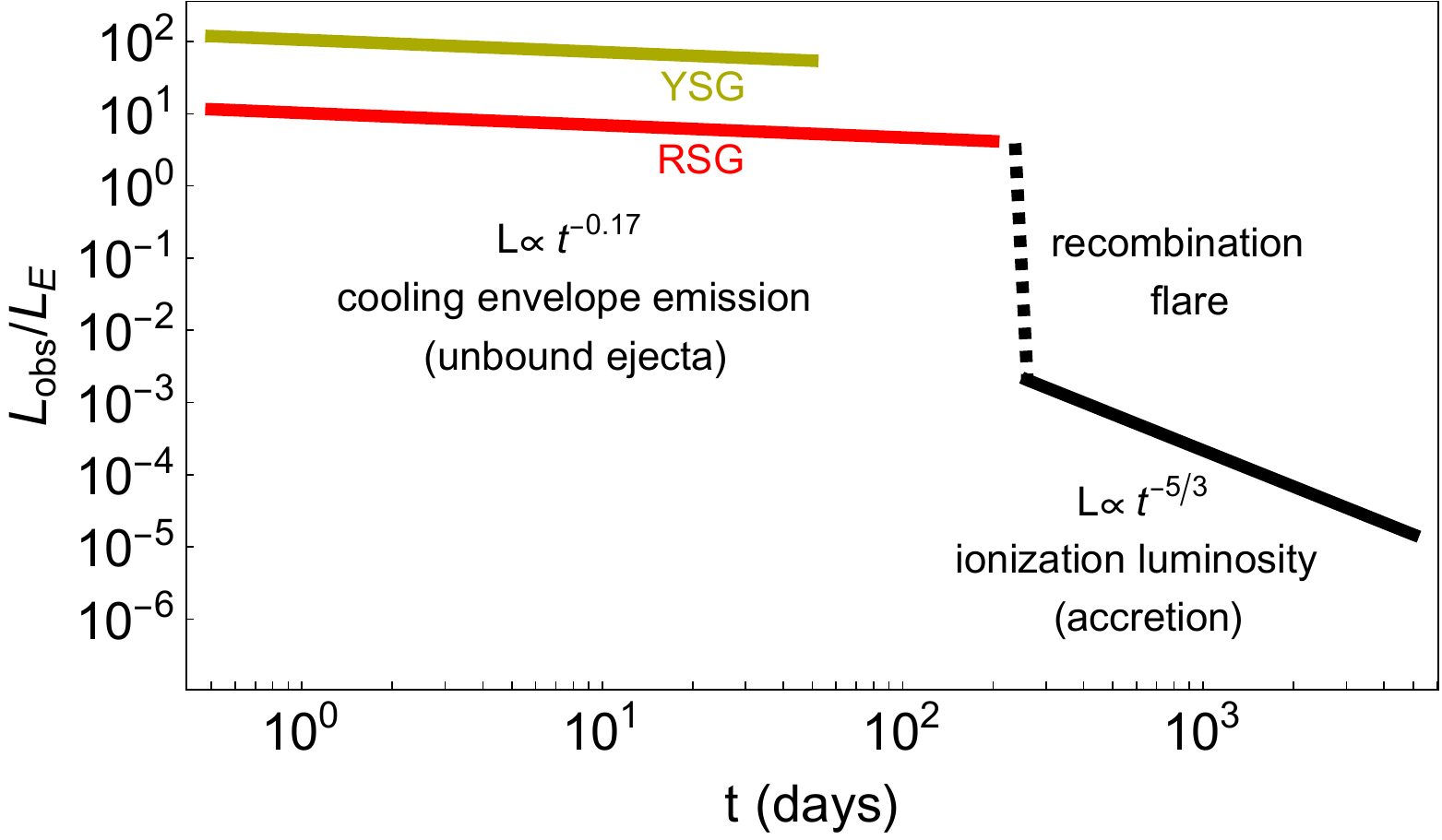}
    \caption{The expected light curve of a low energy core-collapse explosion leading to black hole formation, computed for the properties of our H-rich red supergiant (RSG) and H-poor yellow supergiant (YSG) MESA progenitor models; we assume an explosion energy of $10^{48}$ erg and a black hole mass of $M = 5.6 M_\odot$. The dashed lines reflect the fact that the time and amplitude of the recombination flare depend on the explosion energy and ejecta mass. The cooling envelope luminosity from the unbound ejecta assumes an ejecta mass of $0.5 M_\odot$ (YSG) and $6.5 M_\odot$ (RSG). Altering the explosion energy will primarily affect the luminosity during the  cooling envelope phase but will leave the accretion luminosity at later times almost unchanged.    These calculations are for spherical collapse, explosion, and fallback accretion and can be modified by accretion of the turbulent convective envelope, particularly in RSG models (see text for a detailed discussion).}
    \label{fig:L_t_MESA}
\end{figure*}

The early time luminosity of low energy explosions is dominated by the dynamics of the unbound portion of the envelope, which are similar to those of core-collapse supernovae. The observed signature can therefore be estimated using the cooling envelope model of \cite{Nakar2010}. According to their Eq 29, the cooling envelope luminosity scales as $L_{CE} = 2 \times 10^{42} \text{ erg s}^{-1} M_{ej,15}^{-0.87} R_{500} E_{51}^{0.96} t_d^{-0.17}$, where $R_{500} = R/500 R_\odot$ is the radius from which the shock breaks out, $E_{51} = E/10^{51} \text{erg}$, $M_{ej,15} = M_{ej}/15 M_\odot$ where $M_{ej}$ is the mass of the ejecta, and $t_d = t/day$. Note that these calculations assume radiation pressure dominated ejecta and neglect H recombination energy and so do not apply to energies $\lesssim 10^{48}$ erg.
From the dynamics of the shell satisfying $\tau = c/v$, we find that the shock cooling emission from a $10^{48}$ erg explosion will last $\simeq 50$ days for $M_{ej} = 0.5 M_\odot$ (YSG) and $\simeq 360$d for $M_{ej} = 7 M_\odot$ (RSG). As shown in Fig \ref{fig:L_t_MESA}, the luminosity is initially super-Eddington and decreases slowly with time. This emission decreases abruptly once the recombination wave passes through the bulk of the unbound and marginally bound ejecta (as in II-P supernovae).

The cooling envelope luminosity from the unbound ejecta will be followed by a phase of accretion powered luminosity. Since the duration of the cooling envelope phase is comparable to the upper limit on the recombination time for our progenitors, $\simeq 260 \text{d} (M/5 M_\odot)^{-1/2}$ (see Eq \ref{eq:trec}), we focus on the late time emission from the neutral inflow that obeys $L_{ion} \propto \dot{m}$. Setting $t_0 = r_0/v_0$ and $v_0 = \sqrt{2GM/r_0}$ in Eq \eqref{eq:MdotChevalier} for $\dot{m}$, one obtains $\dot{m} \propto \rho_0 r_0^4 M^{-4/3} t^{-5/3}$. In the inner, non convective zones of RSG and YSG envelopes, $\rho_0 \simeq 0.1 (r/R_\odot)^{-4}$, so that at a given time both $\dot{m}$ and $L_{ion,r}$ depend only on the mass of the black hole. As a result, the late time luminosity according to Eq \eqref{eq:L_ion_r} will be
\begin{equation}
    L_{ion,r}^{Y/RSG}/L_E \simeq 10^{-3} \left(\frac{M}{5 M_\odot}\right)^{-4/3} \left(\frac{t}{1 \text{yr}}\right)^{-5/3}\,.
\end{equation}
Since the enclosed mass (i.e., the black hole mass $M$ formed) is effectively constant for $E_b$ in the range $\sim 10^{45} - 10^{49}$ erg, the above coincidence creates a degeneracy in the late-time observed properties of fallback accretion from RSG (or YSG) progenitors of the same mass with different explosion energies. 
Naïvely, this means that if the time since the explosion is known, late-time observations of the bolometric luminosity can be used to infer the mass of the newly formed black hole. A prediction for the light curve from a $10^{48}$ erg explosion is shown in Fig \ref{fig:L_t_MESA}.   We stress that this prediction is for a spherical explosion and for spherical fallback accretion.



The two most promising candidate failed SNe are the transient in NGC 6946 \citep{Adams2017, Kochanek2024} and that in M31 \citep{De2024}.   In both cases there is a fading source at the previous location of the massive YSG with a luminosity of several $10^4 L_\odot$ after $\sim 5$ years, qualitatively consistent with models in which the YSG collapsed to form a black hole, with fallback accretion powering the late fading emission.    However, the calculations here show that spherical fallback accretion onto a black hole produces a luminosity much less than that observed in both of these events, and should be undetectable (Fig. \ref{fig:L_fallback}).   Our analytic results are consistent with the numerical simulations of \citet{Zampieri1998}, who simulated the collapse of a massive star to form a black hole, and its radiative signature.   How can we reconcile these results with the comparatively luminous fading sources observed by \citet{Kochanek2024} and \citet{De2024}?   We suggest that the most likely explanation (if indeed these sources are failed SNe) is that the collapse and subsequent weak transient cannot be modelled using the spherically symmetric treatment of, e.g., \citet{Lovegrove2013,Fernandez2018}.   Instead, the random convective motions in the supergiant envelope provide both a source of additional dissipation in the inflow (random kinetic energy) and random angular momentum that leads to an accretion disk around the newly formed black hole \citep{Gilkis2014,Quataert2019,Antoni2023}. This can substantially decrease the fallback accretion rate relative to spherically symmetric predictions (because of outflows associated with super-Eddington accretion) while at the same time substantially increasing the luminosity at a given accretion rate relative to that produced by spherical super-Eddington accretion with no internal dissipation.   \citet{De2024} estimated that the black hole accretion rate could be suppressed sufficiently to fall below Eddington after a few years, plausibly consistent with the fading sources observed in NGC 6946 and M31 (see their Fig 4b).  More detailed hydrodynamical simulations to assess this would be very valuable. 
Another key question is how the turbulent convective flows evolve during the expansion to the turnaround radius and the subsequent fallback accretion.    \citet{Gilkis2014,Quataert2019,Antoni2023} showed that the collapse of the supergiant envelope powers rotating accretion flows and outflows at early times but it is not clear if this remains true at later times during fallback accretion (when the turbulent flows in the progenitor envelope have had time to dissipate and/or evolve `adiabatically'; \citealt{Robertson2012}).   



\section{Summary}\label{sec:Summary}
We have studied the effect of photon diffusion on super-Eddington accretion onto a black hole, in the regime where gas pressure is negligible. The first part of the paper is dedicated to a theoretical investigation of the equations, whose solution is used in the second part to infer the observational signatures of fallback accretion in core-collapse supernovae. Below is a summary of our main results.
\begin{enumerate}
    \item We obtain a single ODE for the dimensionless parameter $\Psi$, whose derivative is proportional to the diffusive luminosity, and find that it admits two types of solutions. The first type is isothermal at large radii; beyond the trapping radius, the diffusive and advective luminosities are comparable, and the rest frame luminosity, defined as $L_d-L_a$, increases radially as $r^{1/2}$ beyond the trapping radius. This increase is a result of the sub-dominant departure of the flow from the free-fall limit, owing to the dynamical effect of the constant pressure profile. The second branch is characterized by a vanishing temperature at infinity and a constant diffusive luminosity beyond the trapping radius. Using boundary layer theory, we obtain an accurate analytic solution for the isothermal branch and an approximate, leading order solution for the power-law branch. For a fully ionized accretion flow, the rest frame luminosity at the photosphere, where $r_{ph} \simeq \dot{m}^2 \times r_S$, scales with $\dot{m}$ as follows:
    \begin{equation*}
        L_{obs} \propto \begin{cases}
            \dot{m}^2 & \,, ~~\text{isothermal branch}\\ \dot{m} & \,,~~ \text{power-law branch}
        \end{cases}\,.
    \end{equation*}
    The rest-frame luminosity at the photosphere is suppressed by a factor of $\dot{m}$ with respect to $L_d$, which is significantly smaller than what a naïve estimate of the photospheric luminosity would give.
    
    The exact location of the trapping radius in the two types of solutions, defined as the point where $L_a = L_d$, is found to be
        \begin{equation*}
        r_{tr} = r_S \times\begin{cases}
            3 \dot{m} & \,, ~~\text{isothermal branch}\\ \dot{m}/2 & \,,~~ \text{power-law branch}
        \end{cases}\,.
    \end{equation*}

    Since the solutions are obtained in the free-fall limit, the adiabatic sonic point (or alternatively, the Bondi radius) does not play a role in the dynamics of the flow inside the photosphere. Nevertheless, an analogous role is played by the trapping radius in the isothermal branch. Beyond the trapping radius the flow becomes isothermal, just like in the adiabatic problem beyond the Bondi radius.
    
    \item A natural application of this work is to fallback accretion onto black holes during core-collapse supernovae.  This is particularly interesting in low energy explosions (nominally `failed' supernovae) where the fallback accretion rates are likely high.   
    The accretion rate is known to scale with time as $\dot{m}\propto t^{-5/3}$ in the steady state limit. The balance of compressional heating with cooling processes outside the photosphere imposes a photospheric temperature of $T \lesssim 10^4$ K, close to or below the recombination temperature of hydrogen. Under the (probably incorrect) assumption that this balance can maintain a fully ionized photosphere, we find that the observed luminosity is given by (see Fig. \ref{fig:L_fallback}):
    \begin{equation*}
                L_{obs}\simeq L_E\times\begin{cases}
            (3\times 10^{-13} M_{10} T_4^4) ~\dot{m}^2 & \,, ~\Delta L_{ph}>L_c\\ (2\times 10^{-9} T_4)~  \dot{m} & \,,~ L_c>\Delta L_{ph} \\ \propto\dot{m}^{5/3}& \,,~ p_{g,ph}>p_{r,ph}
        \end{cases}\,,
    \end{equation*}
    where $\Delta L_{ph}$ is the rest-frame luminosity at the photosphere, $L_c$ is the luminosity emitted in the optically thin region due to cooling of gas internal energy, and $p_{g,ph},\, p_{r,ph}$ are the gas and radiation pressure at the photosphere, respectively.
    
    
    \item 
    As the fallback material expands out towards its turnaround radius, a recombination wave will quickly sweep through the flow, producing  a flare of radiation as the internal energy of the gas is rapidly released. Once the local velocity of the gas becomes comparable to the inward propagation velocity of the recombination front, neutral gas that has previously recombined gets ionized as it is compressionally heated.  This modifies the luminosity relative to the assumption of fully ionized gas:         \begin{equation*}
                L_{ion} \simeq L_E \times \begin{cases}
            10^{-7} \times\dot{m} & \,, ~~ p_r>p_g\\ 10^{-8} M_{10}^{-1/3} \times \dot{m}^{4/3}& \,, ~~ p_r<p_g
        \end{cases}\,,
    \end{equation*}
    where the inequalities refer to the radiation and gas pressure at the ionization front.   

    The calculations here assume that the fallback accretion does not cool below $\sim 5000$ K.   However, it is possible that dust can form in the expanding gas prior to its turnaround at apocenter (M31-2014-DS1 \citep{De2024} and N6946-BH1 \citep{Kochanek2024} report observational signatures of dust, though this may be dominated by unbound gas).  We have not included this in our calculations, but it would be interesting to do so in future work.

    \item The analysis in this work was restricted to inviscid flows, in which the sole contribution to the luminosity comes from compressional work. In actual systems, reconnection of magnetic field lines and dissipation of turbulent motions may serve as additional energy sources.  Indeed, we argue that this is necessary to explain the fading sources associated with the failed SNe candidates M31-2014-DS1 \citep{De2024} and N6946-BH1 \citep{Adams2017}:  the spherical fallback accretion luminosities we find are significantly smaller than the observed late-time luminosities of these sources (see Fig. \ref{fig:L_t_MESA}; the same is true in the radiation simulations of, e.g., \citealt{Zampieri1998}).    \citet{Gilkis2014,Antoni2023} show that infall of the turbulent convective envelope in a RSG leads to centrifugally supported gas due to the large random angular momentum associated with the turbulent convection.   Dissipation of these turbulent motions and the release of gravitational energy by accretion can then produce accretion luminosities of order or greater than the Eddington luminosity, as is likely necessary to explain M31-2014-DS1 and N6946-BH1.   An interesting, and unsolved, problem is how the turbulence evolves at late times in fallback accretion, when the infalling gas has expanded out to large radii before reaccreting.  If most of the energy dissipates prior to the gas reaching its turnaround radius, the turbulence will not be a plausible mechanism for enhancing the late time accretion luminosity in fallback accretion.  On the other hand, \cite{Robertson2012} show that random bulk velocities of turbulent eddies can increase (decrease) during contraction (expansion) of a fluid, assuming that the eddies' turnover time is long compared to the contraction time.  If the turbulent motions in RSG fallback accretion remain near turnaround they may thus be further amplified to become dynamically important during the subsequent fallback accretion.

    A prediction of our model is that if the core-collapse of a massive slowly rotating Wolf-Rayet star or blue supergiant does not produce an energetic explosion, the resulting fallback accretion onto the black hole is likely to be very faint because the assumption of spherical symmetry is a much better approximation for these progenitors.  The star would truly `disappear.'



\end{enumerate}

\section*{Acknowledgements}
We thank Andrea Antoni for useful conversations and for providing the MESA models used in this work.   This research benefited from interactions at workshops funded by the Gordon and Betty Moore Foundation through grant GBMF5076. TF is grateful to the CCPP department at NYU for hosting her as a visiting scholar.

\begin{appendix}
\section{Outer boundary layer solution}\label{App:OuterBLSol}
We seek a solution to Eq \eqref{eq:ChiODELinearized} that describes the flow outside the boundary layers at $x=1$ and $x = x_{bf}$. We assume a solution in the form of a perturbation series in $\delta$:
\begin{equation}
    \chi_{out} = \sum _{n=0}^{\infty}\chi_n \cdot \delta^n\,.
\end{equation}
Plugging the expansion into Eq \eqref{eq:ChiODELinearized}, we obtain the equations satisfied for each order $n$:
\begin{equation}
    6x(1-x)\chi'_n-(1-4x)\chi_n =
    \begin{cases}
        0 &\,, n=0 \\-4x(1-x)\chi_{n-1}''-2(1-4x)\chi'_{n-1} &\,, n>0
    \end{cases}
\end{equation}
The equations can be solved analytically; the leading order solution is found to be
\begin{equation}
    \chi_0 = B_1 x^{1/6}\sqrt{1-x}\,,
\end{equation}
which is used to compute the first subleading term
\begin{equation}
    \chi_1 = x^{1/6}\sqrt{1-x} \times\left\{B_2-\frac{B_1}{27}\left[\frac{9}{1-x}+\frac{1}{x}+12\ln\left(\frac{1-x}{x}\right)\right]\right\}\,,
\end{equation}
and so on. The constants $B_1$ and $B_2$ are determined by the conditions on $\chi$ and $\chi'$ at the inner boundary and by matching to the boundary layer solutions.

\section{Evaluation of $\dot{e}$ for isothermal flows}\label{app:edoEvaluation}
The value of $\dot{e}$ can be estimated using the result
\begin{equation}
    \Psi_{ph}' = \frac{\Lambda_a}{\delta(1-6\delta)} \,.
\end{equation}
The change in $\Psi$ over a scaleheight $\Delta x = \delta^2$ is
\begin{equation}
    \Delta \Psi_{ph} \simeq \frac{\delta \Lambda_{a,ph}}{(1-6\delta)}\,,
\end{equation}
which is equal to $\Delta\Lambda_{ph}$. Therefore
\begin{equation}
    \dot{e} =\Delta\Lambda+\frac{\Psi_{ph}}{2} = \frac{ \delta\Lambda_{a,ph}}{1-6\delta}+ \frac{\Psi_{ph}}{2} = \Delta\Psi_{ph}+\frac{\Psi_{ph}}{2}.
\end{equation}
At most, $\Delta\Psi_{ph} = \Psi_{ph}$, which is the case if the initial value of $\Psi_{ph}$ is negligible compared to $\Delta \Psi_{ph}$. Therefore, the value of $\dot{e}$ lies in the range
\begin{equation}
    \frac{\Psi_{ph}}{2}<\dot{e}<\frac{3}{2}\Psi_{ph}\,.
\end{equation}
\end{appendix}
\bibliography{main}{}
\bibliographystyle{aasjournal}
\end{document}